\PassOptionsToPackage{table}{xcolor}
\documentclass[acmsmall]{acmart}
\usepackage[resetlabels]{multibib}
\newcites{game}{Ludography}
\newcommand{\citegameprefix}{G}
\usepackage{xparse}
\let\origcitegame\citegame
\RenewDocumentCommand{\citegame}{O{} O{} m}{%
  \renewcommand{\citenumfont}[1]{\citegameprefix##1}%  
  \origcitegame[#1][#2]{#3}%
  \renewcommand{\citenumfont}[1]{##1}%
}
\usepackage{booktabs}
\usepackage{tabularx}
\usepackage{caption}
\usepackage{multirow}
\usepackage{array}
\usepackage{tikz}
\usetikzlibrary{arrows.meta}
\usepackage[table]{xcolor}
\definecolor{tablepurple}{HTML}{B0E0E6}
\AtBeginDocument{%
  }

\setcopyright{cc}
\setcctype{by}
\acmJournal{PACMHCI}
\acmYear{2026} \acmVolume{10} \acmNumber{7} \acmArticle{GAMES061}
\acmMonth{11} \acmDOI{10.1145/3831313}
\acmISBN{978-1-4503-XXXX-X/2018/06}

\begin{document}

\title{SAVEstate: A Method for Documenting Player Reflection in Digital Games}


\author{Nisha Devasia}
\authornote{A portion of this work was conducted in the author’s capacity as a researcher at foundry10.}
\affiliation{%
  \department{Human Centered Design \& Engineering}
  \institution{University of Washington}
  \city{Seattle}
  \country{USA}}
\email{ndevasia@uw.edu}

\author{Michele Newman}
\affiliation{%
  \department{Information School}
  \institution{University of Washington}
  \city{Seattle}
  \country{USA}}
\email{mnewman@uw.edu}

\author{Safinah Ali}
\affiliation{%
  \department{Administration Leadership and Technology}
  \institution{NYU Steinhardt}
  \city{New York City}
  \country{USA}}
\email{sa9140@nyu.edu}

\author{Julie A. Kientz}
\affiliation{%
  \department{Human Centered Design \& Engineering}
  \institution{University of Washington}
  \city{Seattle}
  \country{USA}}
\email{jkientz@uw.edu}

\author{Jin Ha Lee}
\authornotemark[1] % Reuses the same footnote marker from the first author
\affiliation{%
  \department{Information School}
  \institution{University of Washington}
  \city{Seattle}
  \country{USA}}
\email{jinhalee@uw.edu}

% \renewcommand{\shortauthors}{Devasia et al.}

\begin{abstract}
    In recent years, interest in eudaimonic player experiences (PX) - concerning reflection, meaning-making, and personal growth - has increased. However, most games user research methods are not well-suited to study eudaimonic PX, as they have been developed to evaluate features of hedonic PX, such as flow, immersion, and playability. To more deeply explore eudaimonic PX, we require methods that can 1) investigate how moment-to-moment PX shapes player reflection and 2) explore how players reengage and reinterpret their experiences longitudinally. In this paper, we present \textit{SAVEstate}, a method that uses documentation as a means of studying player reflection. SAVEstate consists of two phases: reflection-in-action and reflection-on-action, which allow researchers to probe players' in-situ reflections and how they reengage with their gameplay, respectively. Using SAVEstate, we were able to observe in-situ meaning-making and connect it to post-game reflection-on-action, and view synchronous sensemaking across multiple participants. We also developed an open-source desktop application for researchers to adapt in their own SAVEstate deployments. We discuss implications for how researchers might use SAVEstate to conduct future research in meaningful PX. 
\end{abstract}

\begin{CCSXML}
<ccs2012>
<concept>
<concept_id>10003120.10003121.10003122</concept_id>
<concept_desc>Human-centered computing~HCI design and evaluation methods</concept_desc>
<concept_significance>500</concept_significance>
</concept>
<concept>
<concept_id>10010405.10010497.10010510.10010513</concept_id>
<concept_desc>Applied computing~Annotation</concept_desc>
<concept_significance>100</concept_significance>
</concept>
<concept>
<concept_id>10010405.10010476.10011187.10011190</concept_id>
<concept_desc>Applied computing~Computer games</concept_desc>
<concept_significance>500</concept_significance>
</concept>
<concept>
<concept_id>10003120.10003121.10003129</concept_id>
<concept_desc>Human-centered computing~Interactive systems and tools</concept_desc>
<concept_significance>300</concept_significance>
</concept>
</ccs2012>
\end{CCSXML}

\ccsdesc[500]{Human-centered computing~HCI design and evaluation methods}
\ccsdesc[100]{Applied computing~Annotation}
\ccsdesc[500]{Applied computing~Computer games}
\ccsdesc[300]{Human-centered computing~Interactive systems and tools}

\keywords{method, artifact, document theory, reflection, eudaimonia, meaningful player experience}

\received{February 2026}
\received[revised]{June 2026}
\received[accepted]{July 2026}

\maketitle

\section{Introduction}
The field of Games User Research (GUR) has commonly utilized data-driven approaches to ``ensure the optimal quality of usability and user experience in video games'' \cite{Drachen_Mirza-Babaei_Nacke_2018}. Traditionally, GUR uses three main types of information: user interview insights, telemetry data, and biometrics \cite{desurvire2013methods}. The methodologies used to collect these data have competing strengths and weaknesses, depending on varying levels of ecological validity, whether they provide quantitative or qualitative insights, and whether they collect data in-situ or retrospectively. 

Historically, many of these methods were created to study aspects of player behavior associated with the \textit{hedonic} paradigm, such as flow, engagement, and positive affect \cite{Cole_Gillies_2022, mekler2014systematic}. For example, telemetry (the remote tracking of in-game behavioral data) is excellent for identifying friction points in level design where players struggle to navigate \cite{sifa2018profiling}, and biometric markers like galvanic skill response (GSR) can effectively measure the immediate arousal generated by gameplay \cite{mandryk2006continuous}. In recent years, however, we have seen a significant surge of interest in \textit{eudaimonic} player experiences (PX)—those that concern internal processes such as reflection, meaning-making, and personal growth \cite{daneels2021eudaimonic, Possler_2024, devasia2025would}. Unlike hedonic enjoyment, eudaimonic appreciation often arises from moments of discomfort, moral ambiguity, or complex narrative contemplation that do not necessarily manifest as fun in the traditional sense \cite{Bopp_Mekler_Opwis_2016, Possler_Daneels_Bowman_2024}. Because these experiences are deeply personal and contextual, qualitative methods are necessary to probe them. However, as they often unfold  and evolve over longer temporal scales \cite{devasia2025would, antognozziludic, Mekler_Iacovides_Bopp_2018}, they are difficult to capture using methods designed for reflection-in-action \cite{schon2017reflective} such as think-aloud protocols, challenging to recount retrospectively \cite{vakeva2024disorientation} in interviews or surveys, and extend beyond the typical duration of a diary study.

An additional challenge is that traditional GUR methods are meant to capture instances of play and not necessarily of \textit{replay}. Replay is the ``critical phase where players revisit, reinterpret, and socially integrate their gameplay experiences, creating deeper connections between ludic engagement and broader contexts of life and meaning'' \cite{antognozziludic} and is a crucial component of meaning-making. However, it is seldom studied, despite empirical work showing that meaningful effects resulting from eudaimonic gaming experiences take several years to fully manifest \cite{devasia2025would, Mekler_Iacovides_Bopp_2018}. 

In the process of replay, players' reflection-in-action and reflection-on-action reinforce each other over time \cite{Mekler_Iacovides_Bopp_2018}. \citet{antognozziludic}'s temporal architecture of play-loops posits that transformative and critical reflection occurs over several recursive cycles of replay. This begins from \textit{micro-loops} (constituting the immediate experience of gameplay) and building, over years, to \textit{macro-loops} (operating at cultural scales where design practices, play culture, and broader social meanings co-evolve). Thus, understanding the full evolution of the eudaimonic PX from micro-loops to macro-loops requires GUR methods that are well-suited to encompass the process of replay in three ways: ``(1) how moment-to-moment player experience shapes reflection-on-action, (2) how this develops outside of play over time, and (3) how it may feed back into play and affect subsequent reflection-in-action'' \cite{Mekler_Iacovides_Bopp_2018}.

% In previous work, we encountered the difficulty of drawing conclusions about how higher-level reflection occurred through gameplay without access to a player's thought processes both during gameplay and during their interactions with the broader game habitus. Beyond the gameplay experience itself, players engage in participatory practices that leave behind artifacts that researchers can study, such as Reddit threads \cite{Vakeva_Hamalainen_Lindqvist_2025}, Steam reviews \cite{Phillips_Klarkowski_Frommel_Gutwin_Mandryk_2021}, and Twitch streams \cite{de2020live} - to investigate how and why meaning forms through player experiences.

To understand replay and how we might develop methods for studying it, we can draw inspiration from \textit{document theory} \cite{buckland1991information, frohmann2004documentation, briet2006documentation}. Document theory treats documentation as a process through which information is generated and considered shareable \cite{frohmann2004documentation}. Applied to PX research, a document can be viewed as a reflection upon what aspects of a gameplay experience are meaningful and worth preserving through engaging in cognitive, social, and material means \cite{buckland1991information, buckland1997document, briet2006documentation}. The process of documentation may give us a way to consider and better capture replay. Indeed, researchers have already investigated how artifacts created by participatory cultures — such as Reddit threads \cite{Vakeva_Hamalainen_Lindqvist_2025}, Steam reviews \cite{Phillips_Klarkowski_Frommel_Gutwin_Mandryk_2021}, and Twitch streams \cite{de2020live} — spread and grow the meaning of a game \cite{jenkins2008convergence}. These artifacts are documents in that they have evidentiary value for both players and researchers \cite{frohmann2004documentation,skold2018understanding, jenkins2008convergence}: players can reflect on them and share them with others, and they provide researchers with essential insight into the post-gameplay meaning-making that occurs within affinity spaces \cite{jenkins2008convergence}. Namely, these artifacts help researchers understand ``(2) how [reflection-on-action] develops outside of play over time'' \cite{Mekler_Iacovides_Bopp_2018} on the scale of macro-loops. However, we lack similarly useful artifacts that address ``(1) how moment-to-moment player experience shapes reflection-on-action'' and ``(3) how replay may affect subsequent reflection-in-action'' \cite{Mekler_Iacovides_Bopp_2018}, due to a lack of appropriate GUR methods for generating them. 

Hence, we posed the following research questions: 

\begin{itemize}
    \item How can we leverage the affordances of documentary practices to aid players in reflecting in and on their own gameplay?
    \item How might these practices enable researchers to expand the current boundaries of eudaimonic PX study?
\end{itemize}

We addressed these research questions through the development of the Situated Annotation for Virtual Experience States method (referred to as SAVEstate), which we present in this work. SAVEstate utilizes the documentary practice of annotation \cite{marshall1997annotation} (afforded by a digital annotation tool) to create the critical distance \cite{Khaled_2018} necessary to prompt player reflection during gameplay. At its core, SAVEstate is heavily grounded in investigating "how moment-to-moment player experience shapes reflection-on-action" and "how it may feed back into play and affect subsequent reflection-in-action" \cite{Mekler_Iacovides_Bopp_2018}. Correspondingly, it consists of two phases: a 'reflection-in-action' phase in which participants stabilize a series of contextual anchors (annotations synchronized with screen recordings of gameplay) into a document. The document can then be reviewed in a 'reflection-on-action' phase as a form of 'replay' to revisit and reinterpret past gaming experiences. The method can easily be scaled to collect and analyze these documents from multiple players in a study. 

We describe affordances, tensions, and takeaways from using SAVEstate in three case studies: 1) a 'game book club' conducted remotely with 8 high school students, 2) collecting personal reflections from 17 adults while gaming, and 3) a classroom session with 10 middle schoolers playing and sharing reflections on an educational game. To generate SAVEstate documents at scale, we created a tool called SavePoint, a capture and access application \cite{truong2009ubiquitous} for desktop that enables in-situ annotations for users as they play. We collected valuable insights on how and why players reflected in-game and connected their reflections-in-action to meaning-making and exo-game reflection-on-action. We present implications for future eudaimonic PX research and additionally discuss the value of player-authored data for game preservation more broadly. 

This work contains two primary contributions for the study of play and replay: 1) a methodological contribution for stabilizing reflection-in/on-action during play that benefits both researchers and players, and 2) an artifact contribution that researchers can adapt for their own SAVEstate deployments. In doing so, we hope to make it easier for the CHI PLAY community to advance the field's understanding of how players' in-game reflections can shape meaningful PX and behavior change more broadly. 

\section{Related Work} \label{rw}

\subsection{Qualitative Methods and Games User Research}
A variety of qualitative methods have emerged to study player experience, each bringing distinct strengths to the research landscape. These approaches in game user research (GUR) are designed to illuminate diverse aspects of play, from player behavior and emotional response to post-play reflection and feedback.

A major limitation of many qualitative GUR methods is their reliance on retrospective accounts, which are prone to inaccuracy and memory bias \cite{kahn2014we}. These limitations have previously been discussed for surveys \cite{Bruhlmann_Mekler_2018} and user interviews \cite{Bromley_2018}, both commonly used in GUR. Stimulated recall (SR) protocols address memory biases by prompting participants to retrospectively narrate their thoughts while reviewing a recording of their own gameplay \cite{pitkanen2015studying, lyle2003stimulated}; however, SR narratives may not always be representative of the conscious or unconscious cognition taking place at the time of the videotaped episode \cite{lyle2003stimulated, wilcox1998constructing}. User reviews can provide some insight as to how players interacted with the game in an ecologically valid setting, but user attributes are generally unknown, and reviews are generally retrospective and of mixed quality \cite{bruckner2020play}.

Furthermore, scalability and participant burden further limit current qualitative methods that attempt to collect in-situ PX data. It is difficult to collect user interview data at scale, so survey methodologies are often utilized as an alternative. Think-aloud protocols can feel unnatural for participants \cite{Knoll_2018} and can be time-consuming for researchers \cite{hoonhout2022let}. Some researchers have used the diary method to investigate ``how players thoughts and emotions about a game change over time'' \cite{mekler2014diary}; however, they remain uncommon in GUR contexts as they are considered time-consuming and high risk \cite{hillman2016diary}. Although methods such as systematic self-observation diaries have been used to study eudaimonic PX \cite{Whitby_Iacovides_Deterding_2023}, they can be high burden for participants and researchers alike.

Additionally, ecological validity is often compromised when studies are conducted in laboratory settings rather than natural play environments \cite{Louvel_2018}, as the levels of attention and arousal that players experience in natural environments may be misrepresented \cite{takatalo2011user}. Diary studies offer the closest approximation of ecologically valid reflections in-situ to gameplay, but it can be difficult for players to provide data that includes the immediate context of the gameplay itself. For example, a recent diary study of adolescent gaming habits asked participants to use smartphones for taking photos and videos of their computer or television screen \cite{Munteanu_Mo_Potapov_George_Miller_Singh_2026}. 

Finally, as per \citet{pitkanen2015studying}: ``A particular challenge in the domain of game research is that since many
game researchers are game enthusiasts themselves, researchers might interpret the participant’s statements through the lens of their own experience as game players''. Indeed, current qualitative GUR methods prioritize researcher collection and interpretation of player data. As player interpretation of their own gameplay is a crucial part of eudaimonic player experiences \cite{antognozziludic}, we argue that players should be able to access and reinterpret the rich in-situ qualitative insights that they provide to researchers. 

Together, these limitations point to a consistent gap: current qualitative GUR methods are well-suited to capturing either in-situ behavioral data or retrospective qualitative reflection, but not both. Furthermore, these methods rarely give players a role in interpreting the data they generate. These are gaps that our proposed method aims to address. 

\subsection{Games and Reflection}
Reflection is not a peripheral aspect of play, but a fundamental process through which players make sense of, learn from, and assign meaning to their gameplay \cite{schon2017reflective}. \citet{Khaled_2018} positioned games as``reflection machines'', as through interaction with the game, players are encouraged to form hypotheses about the game world and to consider the relationship between reality and virtual identities \cite{gee2003video}. 

In the last decade, there has been growing interest in reflection as an outcome of game design and play. Games are thought to support multiple forms of reflection; for example, \citet{iacovides2014player} draws on \citet{schon2017reflective} to demonstrate how different player strategies constitute both "reflection-in-action" (occurring during play) and "reflection-on-action" (occurring during breaks in gameplay). \citet{Mekler_Iacovides_Bopp_2018} tie games to \citet{Baumer_2015}'s dimensions of reflective informatics in that games "confront players with puzzling or surprising situations [breakdown], which invite them to plan, experiment [inquiry] and look for new solutions [transformation]". 

Traditional conceptions of player experience (PX) often emphasize flow as the hallmark of optimal gaming \cite{mekler2014systematic}. Flow describes a state of deep immersion where players lose track of time, become less self-conscious, and perform at their peak, resulting in intense enjoyment \cite{csikszentmihalyi1990flow}. However, the flow state and immersion may not actually be conducive to reflection-in-action, as when we are immersed in games, we lack analytical perspective and critical distance \cite{Khaled_2018, Iacovides_Cutting_Beeston_Cecchinato_Mekler_Cairns_2022}.

Empirical studies support this tension. While \citet{Mekler_Iacovides_Bopp_2018} found that transformative and critical reflection \cite{Fleck_Fitzpatrick_2010} are mostly absent from everyday player experience, \citet{Whitby_Iacovides_Deterding_2023} demonstrated that certain in-game events (``endo-game triggers'') can lead to deep, exo-game reflections about players’ own lives or broader societal issues. However, in our previous work, we discussed how behavioral changes resulting from eudaimonic gaming often take years to manifest, making them difficult to study \cite{devasia2025would}. \citet{antognozziludic} temporal architecture of play-loops helps clarify this process by describing how reflection unfolds over different timescales: \textit{micro-loops} (minutes to hours), constituting the immediate phenomenology\footnote{Phenomenology refers to the study of the structures of consciousness and lived experience from the first-person perspective. It seeks to understand how individuals interpret their lived experiences by focusing on perception, memory, and emotion, aiming to describe rather than explain phenomena.}of gameplay; \textit{meso-loops} (days to months), constituting reflection across multiple sessions, as players relate game challenges to personal situations; and finally, \textit{macro-loops} (years to decades), constituting cultural and personal transformation through accumulated play and reflection. In other words, while a single game session over a micro-loop may have little lasting impact, the cumulative effects of many sessions over meso- and macro- loops, each shaped by ongoing reflection and reinforced through social interaction, can fundamentally alter how someone understands themselves \cite{antognozziludic}. Examples of how PX researchers have studied these effects include \citet{Vakeva_Hamalainen_Lindqvist_2025}'s thematic analysis of Reddit posts about \textit{Dark Souls}, which investigates players' reflection-on-action on the scale of meso- and macro- loops. Our previous work discussed how eudaimonic gaming experiences had tangible effects on people's lives on a macro-loop scale through a retrospective survey study \cite{devasia2025would}. \citet{Whitby_Iacovides_Deterding_2023} and \citet{Munteanu_Mo_Potapov_George_Miller_Singh_2026} both employ diary studies with structured prompts intended for micro-loops of play, and conduct semi-structured interviews using participants' diary entries to collect reflection-on-action on a meso-loop scale. 

However, none of these methods are well-suited to collect player reflection across all three timescales of micro-, meso-, and macro-loops. Indeed, recounting how transformation occurred across large temporal scales is difficult for players \cite{vakeva2024disorientation}, as they typically cannot access reflections from micro-loops. Arguably, PX research methods are specifically less well-suited to collecting naturalistic micro-loop reflection; even micro-phenomenological diary approaches such as \citet{Whitby_Iacovides_Deterding_2023}'s do not collect minute-to-minute reflections, and these reflections are not usually coupled with the game context. This risks losing an accurate depiction of a micro-loop as players must reconstruct how they felt minutes to hours prior, often from limited context. What is needed, then, is a method that can stabilize micro-loop reflections alongside their context to allow for accessible reinterpretation at meso- and macro-loop scales. In the following section, we argue that annotation, viewed through the lens of document theory, offers a principled foundation for this.

\subsection{Annotation and Document Theory}
One promising approach to addressing the gap in capturing player reflection is annotation: the act of creating hypertext to mark and comment on moments of interest. Annotation has long been used to capture experience, provide context, and aid memory \cite{hansen2006ubiquitous, marshall1997annotation}. Digital annotation tools now enable users to quickly capture insights that would otherwise not be captured in such contexts (e.g., \cite{van2009note}), enhance engagement with reflection across many settings \cite{colasante2011using, mirriahi2018effects}, and help users sensemake in a variety of contexts, such as web search \cite{hearst2013sewing, kuznetsov2022fuse}, literature review \cite{rachatasumrit2021forsense}, or desktop workspaces \cite{jeuris2014laevo}. In our context, annotation holds value in that it provides connection between observed actions in gameplay and the player's provided reasoning behind them \cite{marshall1997annotation}. 

In games, researchers have experimented with systems that allow players to draw live visual annotations over gameplay streams, primarily for guiding and collaborating with teammates in multiplayer online battle arena (MOBA) games \cite{riegler2014videojot, wuertz2017players, alharthi2018investigating}. However, live annotations are often ephemeral, intended to be consumed and discarded within seconds. Many of these systems demonstrate the utility of live annotation for immediate task performance, but overlook the reflective potential of the data captured.

To fully realize annotation's reflective potential in gaming contexts, we can draw on document theory, which considers how documents function as dynamic records of interpretation, negotiation, and meaning-making \cite{frohmann2004documentation, frohmann2008documentary, lund2009document}. Originating in library and information science, document theory posits that documentation - constituting of people engaging in reflection and generating knowledge - is a process through which information is generated and considered shareable. Crucially, document theory uses the term ``document'' differently from its everyday sense: a document is an object that holds evidentiary value to its creator and can be viewed as a judgment by which the material aspects of the process come to represent an experience and knowledge \cite{buckland1991information}. Material aspects can be physical or digital artifacts (e.g., notes, annotations, or screenshots) that help stabilize and communicate knowledge. 

In \citet{newman2026documentation}'s design framework for documentation in play, the authors outline four mechanisms through which players engage in evidence construction within games: interpretive attention (the process by which players notice, select, and make sense  of cues in the game world, drawing on their own prior knowledge and experiences), social validation (the process by which players collectively negotiate what counts as credible evidence), material inscription (the process by which players stabilize, externalize, and manipulate information through persistent traces within the game), and recursive production (captures the ongoing, iterative interplay between the three prior mechanisms). These mechanisms situate documentary practices in play within an epistemological framework where the practice of documenting is itself the data \cite{frohmann2004documentation}. 

The idea of documentation as data relates to our study of reflection in two primary ways. First, as discussed by \citet{newman2026documentation}, the documentation that players generate can make prior decisions visible and support their reflection upon the game world. Second, as per \citet{makela2018documentation}: ``documentation...functions as conscious reflection [in/on] action''. In other words, providing players with the scaffolding and tools to document their play will simultaneously allow us to study reflection. We believe that this approach remains unaddressed by current GUR methods, and intend to investigate it with our proposed method, SAVEstate. SAVEstate uses annotation as the medium of documentation, and utilizes these annotations as \textit{evidence} of the experience of play \cite{briet2006documentation}. In the following section, we present our methodological approach, outlining how annotation and document theory can be operationalized to capture and empower player reflection.

\section{SAVEstate}

\subsection{Overview}
\begin{figure}
  \centering
  \includegraphics[width=0.5\linewidth]{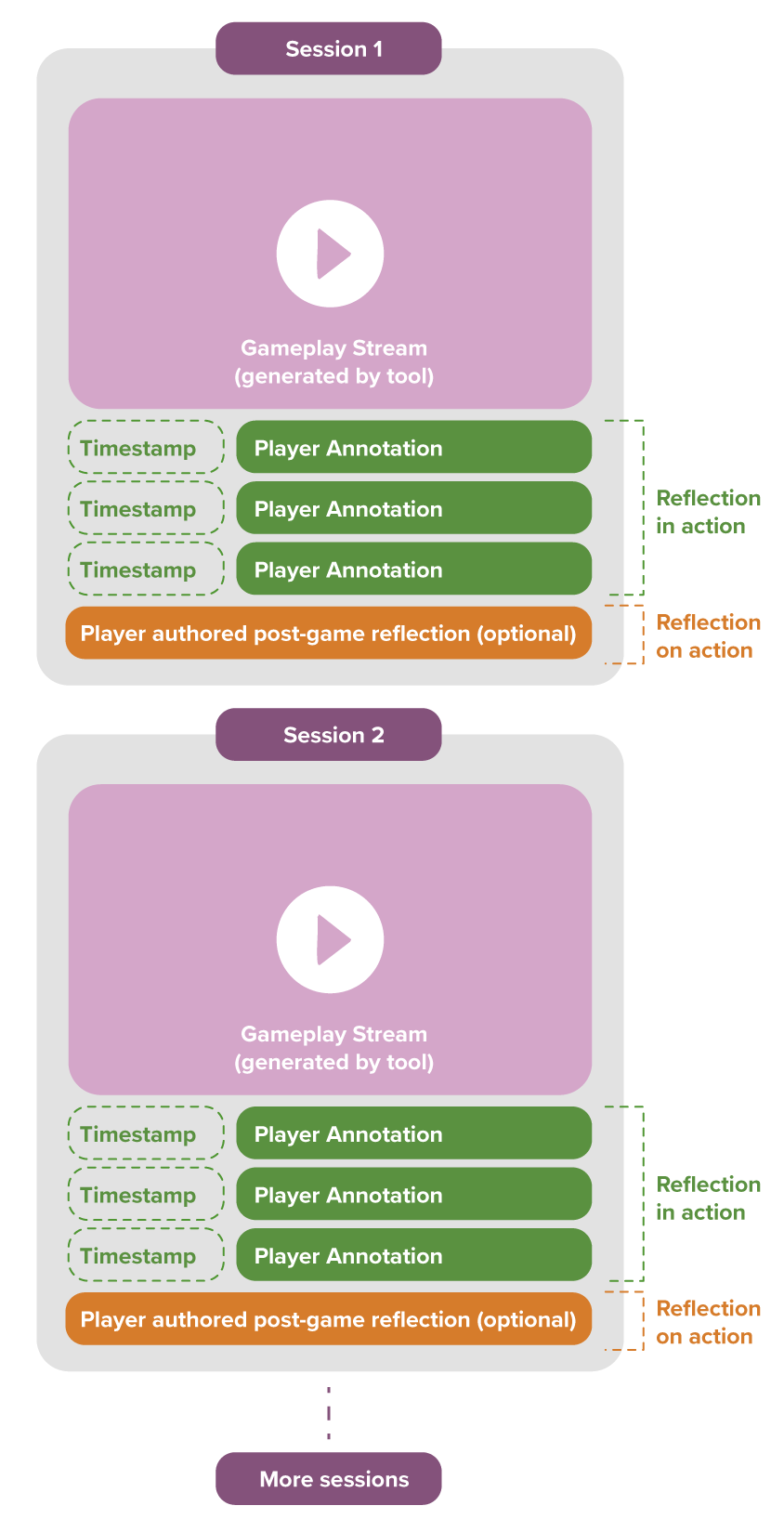}
  \Description{Description of a SAVEstate document}
  \caption{Anatomy of a SAVEstate document, the output of the method. Each document contains multiple sessions, comprised of a set of \textit{contextual anchors} (annotations hyperlinked to the gameplay stream).}
  \label{anatomy}
\end{figure}
SAVEstate (Situated Annotation for Virtual Experience States) is a method for collecting naturalistic player reflection during and after gameplay, operationalized through a digital annotation tool (see Section \ref{SavePoint}). Like experience sampling \cite{hektner2007experience} or stimulated recall \cite{pitkanen2015studying}, SAVEstate is best understood a structured set of procedural decisions (see Section \ref{procedure}) about when and how to elicit reflection, made feasible in naturalistic settings using a purpose-built tool. The method consists of two phases grounded in \citet{schon2017reflective}'s framework of reflective practice. In the reflection-in-action phase, players annotate their gameplay in-situ using a digital tool, producing a set of \textit{contextual anchors} (annotations hyperlinked to a synchronized gameplay recording) that materially couple moment-to-moment reflection with the gameplay context. In the reflection-on-action phase, players and researchers revisit these anchors to ``replay'' \cite{antognozziludic} and reinterpret the captured reflections at any point post-play. The output of SAVEstate is a \textit{document} (see Figure \ref{anatomy}) made up of multiple \textit{sessions}. A session is an annotated gameplay stream, comprising a set of contextual anchors. Because these documents are player-authored and player-accessible, they support the kind of longitudinal reengagement that characterizes meaningful replay by allowing participants to revisit their documents across sessions, editing and reinterpreting their anchors and reflections-on-action as their understanding of the game evolves over meso-loops. These insights can accumulate over macro-loops and can be used to investigate how gameplay influenced higher-order reflection, behavior change, and collective social practices.

SAVEstate is positioned at the intersection of qualitative GUR methods, particularly adaptations of the think-aloud protocol (e.g. \cite{pellicone2022playing}), stimulated recall methods, and diary study methods. Think-aloud protocols capture reflection-in-action but are often lab-based and place cognitive load on participants through verbal narration. Additionally, the outputs of these protocols are typically not made available to the participant. Stimulated recall builds on think-aloud protocols by anchoring retrospective reflection to a gameplay recording, but remains lab-based and researcher-directed. Diary studies excel at capturing reflection-on-action longitudinally in naturalistic settings, but do not collect moment-to-moment reflection-in-action during play; entries are authored without access to the gameplay context itself, which is burdensome for both players and researchers. SAVEstate inherits the in-situ timing of think-aloud, the gameplay-anchored structure of stimulated recall, and the naturalistic, player-authored character of diary studies. We further argue that technical affordances of SAVEstate (Section \ref{SavePoint}) allow players to reflect more naturally by directly annotating their gameplay \cite{hansen2006ubiquitous}, rather than switching mediums to write or record in a diary (e.g. \cite{Munteanu_Mo_Potapov_George_Miller_Singh_2026}). The methodological contribution lies in using the affordances of documentary practices to bridge in-situ and retrospective reflection in naturalistic settings.

SAVEstate aims to create new knowledge by providing a new lens for conducting \textit{practice} games research (concerned with the emergent practices and experiences that occur as a result of interaction with games) \cite{carter2014paradigms}, and inform best practices for future games research that may benefit from qualitative, in-situ, and naturalistic user data. In line with \citet{wobbrock2016research}'s specifications for methodological contributions, we provide three case studies of the method in action to address reproducibility and validation by repeated application. As demonstrated through our development of SavePoint, a desktop tool built to support SAVEstate, our methodology implications are grounded in "the review or development of data-gathering instruments and tools, informing how research data is collected" \cite{van2023implications}.

\subsection{Designing a SAVEstate Study} \label{procedure}

Researchers deploying SAVEstate must make a series of decisions across its two phases. We outline the key considerations for each below, summarized in Table \ref{table:scaffolding-considerations}.

\subsubsection{Phase 1: Reflection-in-Action}
The reflection-in-action phase asks players to annotate their gameplay in-situ, breaking the flow state and creating critical distance from the play experience that allows for reflection \cite{Iacovides_Cutting_Beeston_Cecchinato_Mekler_Cairns_2022}. Thus, the researchers must first consider when to create these breaks through \textit{degrees of scaffolding} (Table \ref{table:scaffolding-considerations}). In a low scaffolding case, participants self-determine when to reflect, and the researchers do not explicitly create breaks past asking participants to annotate their play. For medium scaffolding, the researcher must create the breaks by providing participants with guidelines as to when to reflect. These guidelines can be event-based (e.g., player death, cutscenes, level transitions, item acquisitions), affect-based (tied to emotional valence, such as moments of strong frustration, elation, or surprise), strategy-based (tied to complex problem-solving or decision making moments in games), or reflection-based (tied to connections the player makes with the real world). Finally, high scaffolding requires the researcher to create system-initiated breaks for the participant to reflect at. These be prompted based on pre-programmed logic (e.g., a timer), input reading (e.g., user is prompted to annotate when they have not pressed a button for a certain amount of time), or computer vision events (e.g., detecting certain parts of the game, such as cutscenes). However, researchers must consider using system-initiated annotations might be obtrusive and interrupt flow more than strictly necessary.

Next, researchers must determine the \textit{level of articulation} they desire from participants' reflection-in-action (Table \ref{table:scaffolding-considerations}). All the levels must invoke some amount of critical distance, meaning that even low levels of articulation should require participants to consciously press a button to reflect. In a gaming context, typed reflection has the highest level of articulation, as it requires participants to pause the game and structure their thoughts; however, it is necessarily high user burden and may not be possible in certain genres of games. Oral reflection represents a medium level of articulation, as its cognitive overhead is lower than written reflection \cite{bourdin1994written}, especially while gaming. Finally, simple low articulation hotkey presses could serve as stand-ins for later reflection while simultaneously providing some information to the researcher as to when the participant wanted to reflect.

\subsubsection{Phase 2: Reflection-on-Action}
In the reflection-on-action phase, players revisit the contextual anchors generated during play and decide which serve as useful evidence for their experience \cite{buckland1991information, briet2006documentation}. This phase shares structural similarities with stimulated recall \cite{pitkanen2015studying}, in that both use a gameplay record to prompt retrospective reflection. For this phase, researchers must decide the \textit{level of review} they wish to support. A low level of review simply entails the participants reviewing the document of generated contextual anchors. This is considered the bare minimum of the method, and we recommend prompting it automatically for participants. A medium level of review would require participants to modify, delete, or re-interpret their anchors post-session. This would be especially useful if participants primarily annotated with low levels of articulation, as the review would allow them to clarify to themselves and to the researcher what meaning they intended to convey. At a high level of review, participants are asked to perform synthesis, either individually (writing a post-session summary using their contextual anchors) or socially (e.g., researcher-moderated discussions that rely on the generated documents). 

\begin{table}[h]
\small
\centering
\caption{Considerations for Designing a SAVEstate study}
\label{table:scaffolding-considerations}
\begin{tabular}{@{} >{\raggedright\arraybackslash}p{0.20\linewidth} >{\raggedright\arraybackslash}p{0.15\linewidth} >{\raggedright\arraybackslash}p{0.60\linewidth} @{}}
\toprule
\textbf{Consideration} & \textbf{Phase} & \textbf{Details and Options} \\ \midrule

\textbf{Degree of Scaffolding} & Reflection-in-action & \textbf{Low (User-initiated):} Participant defines significance as emergent to their experience. \newline \textbf{Medium (Researcher-guided):} Guidelines provided based on research questions. \newline \textbf{High (System-initiated):} Prompted via pre-programmed logic, input reading, or computer vision. \\ \midrule

\textbf{Level of Articulation} & Reflection-in-action & \textbf{Low (e.g. hotkey react):} Minimal disruption to flow; serves as stand-in for later review. \newline \textbf{Medium (e.g. voice-to-text):} Supports oral reflection with voice memos that can be transcribed and synced as anchors. \newline \textbf{High (e.g., written):} Use of text window to make typed notes; encourages precise articulation.\\ \midrule

\textbf{Level of Review} & Reflection-on-action & \textbf{Low:} Participants are shown the contextual anchors generated during session, but review is not enforced. \newline \textbf{Medium:} Participants are asked to modify, delete, or re-interpret contextual anchors, and to generate an unstructured post-session summary of their play. \newline \textbf{High:} Participants are asked to modify, delete, or re-interpret contextual anchors. They are given structured reflection questions to answer in a post-session summary. \\ \bottomrule
\end{tabular}
\end{table}

\subsection{Technical Considerations for a SAVEstate Study}
SAVEstate requires custom-built \textit{capture and access applications} (C\&A) to collect contextual anchors (gameplay video and annotations) from the player's typical gaming environment. Capture and access applications focus on seamlessly preserving live experiences for later retrieval \cite{truong2009ubiquitous}. In the context of SAVEstate, such applications would need to blend into the player's natural environment to "capture" reflection-in-action while facilitating subsequent "access" for reflection-on-action. Drawing on metaphors from document theory, C\&A applications can create "an interactive memory or file cabinet" in which the player's "context can be entered through a keyboard and active editor, retained and modified indefinitely, and displayed on demand" \cite{kay1977personal}. Furthermore, players are given agency to curate their annotations in a way that is evidentiary of their experience \cite{buckland1991information}. 

A C\&A application for SAVEstate would need to use platform-native recording and input for the game platform being studied (e.g., desktop, mobile, console). Different game platforms afford different solutions; for example, while most platforms have built-in screen recording functionalities, capture cards or external applications may allow users to have more customizability over their recordings. Input modalities may also vary by platform, e.g., short-form text entry may be difficult on controllers. 

Once the gameplay session is concluded, the raw data must be transformed into a metadata object suitable for player interpretable review. We recommend including, at minimum, the following parameters:
    \begin{itemize}
        \item \textit{Session start [timestamp]}: This records the system time from the instantiation of the screen recording session.
        \item \textit{Annotations [list of {annotation content, timestamp}]}: The timestamp should be recorded relative to the beginning of the recording so that the session recording can be indexed into at the moment an annotation was made. 
        \item \textit{Screen recording [video file]}: File formats include .mp4 or .mkv as examples, but can be chosen by the researcher depending on preferences.
        \item \textit{Game name [string]}: This could be used to sort sessions by game played.
        \item \textit{Date of play [timestamp]}: The system timestamp can be converted into a human readable date to chronologically sort annotation sessions.
    \end{itemize}

Finally, data needs to be uploaded from the user's machine to a centralized database for researcher access. Automated uploads are convenient for both players and researchers, but user-initiated uploads allow participants to edit and reflect on their thoughts before presenting them to the researcher. 

\section{SavePoint: A Desktop Tool for Supporting SAVEstate} \label{SavePoint}
\begin{figure}
  \centering
  \includegraphics[width=0.7\linewidth]{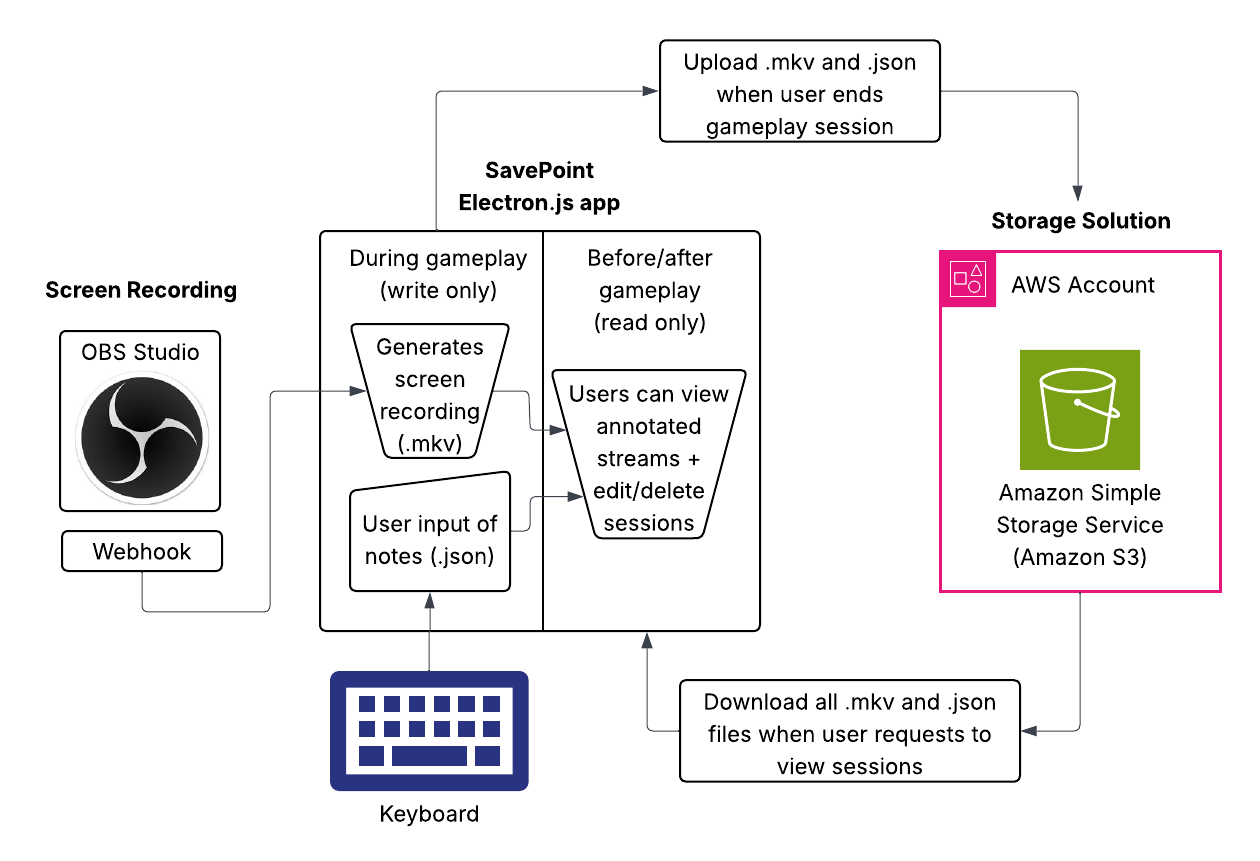}
  \Description{System diagram of SavePoint}
  \caption{A diagram of the SavePoint system showing its main components: the screen recording application (currently configured with Open Broadcast Studio webhook), the Electron.js user interface, and the storage solution (currently configured with AWS S3).}
  \label{system}
\end{figure}

\begin{figure}
  \centering
  \includegraphics[width=0.9\linewidth]{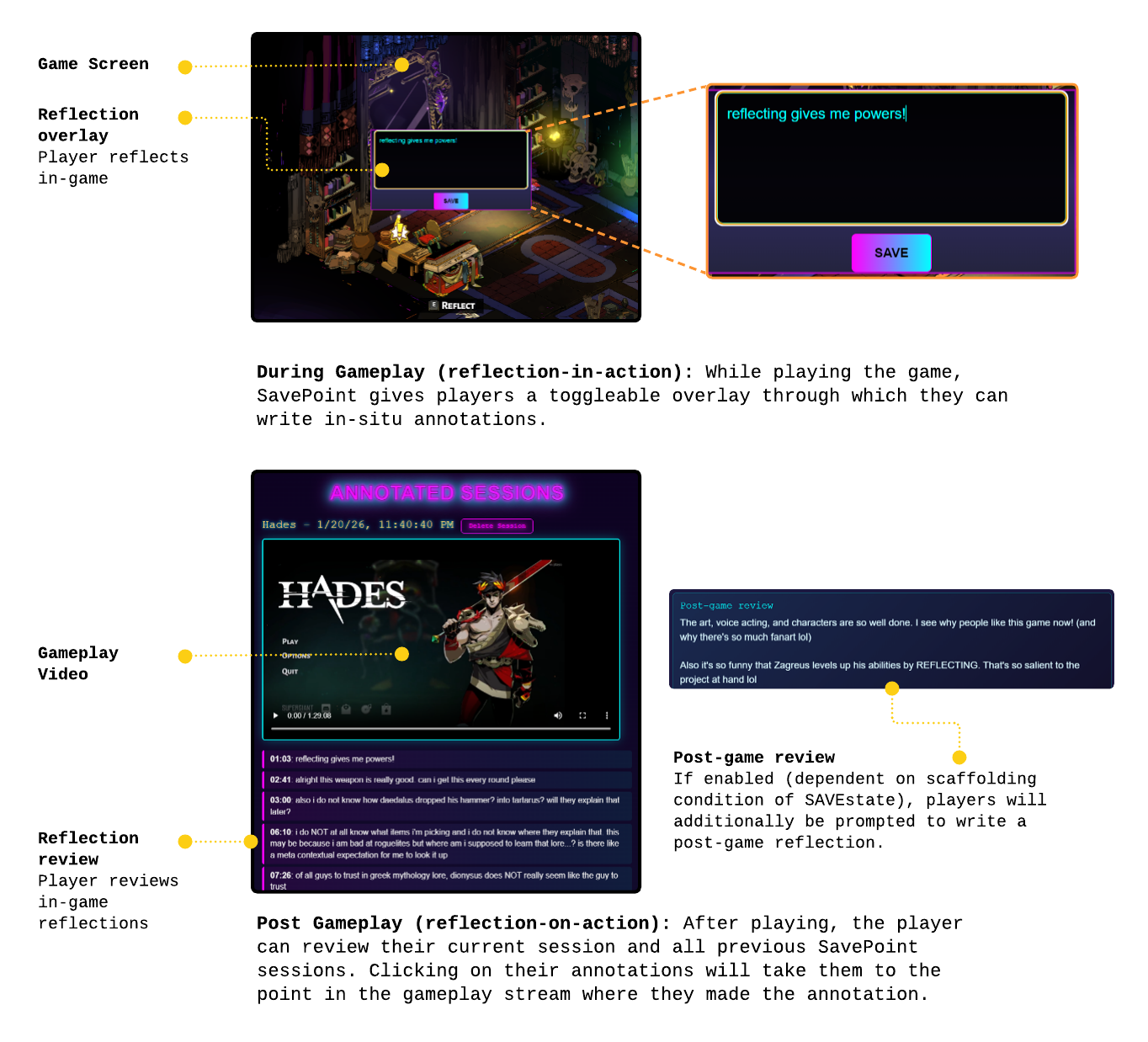}
  \Description{User flow of SavePoint}
  \caption{The core interactions (reflection-in-action and reflection-on-action) within the SavePoint application.}
  \label{flow}
\end{figure}

To operationalize the considerations of SAVEstate, we built SavePoint\footnote{https://github.com/ndevasia/SavePoint}, a C\&A desktop application for Windows and MacOS (see Figure \ref{system}), to collect data in our SAVEstate deployments. SavePoint serves as a metaphorical ``file cabinet'' \cite{kay1977personal} for the documents generated by the SAVEstate method. It eases the burden of data entry for participants by maintaining indexical representation, allowing players to annotate without switching out of the gameplay window (see Figure \ref{flow}). It also simplifies data collection for researchers with automated uploads to a centralized cloud database. 

In designing SavePoint, we sought to support the compilation of contextual anchors into reusable documents while striking the right balance between critical distance and immersion. To make an annotation, the user must necessarily break immersion to press a combination of buttons. However, by linking annotations to the gameplay stream, players can focus on the construction of a document of their interpretations without the burden of describing the game state. As noted by \citet{hansen2006ubiquitous}: "the more precise the annotation is anchored into a context, the easier it is for the annotator to use deictic expressions and write shorter messages relying on semiotic indexes." Thus, players can use shorter messages that can be edited post-play in their documents and not significantly break their immersion. 

In terms of SAVEstate's study design considerations, SavePoint supported the following:
\begin{itemize}
    \item \textbf{Degree of instructional scaffolding}: SavePoint supported \textit{low} and \textit{medium} levels of scaffolding for annotations. We did not utilize system-initiated triggers in our testing of the prototype.
    \item \textbf{Level of articulation}: SavePoint explicitly supported \textit{high} levels of articulation in the form of a textbox that players could annotate in while playing. We also provided emote reacts as a \textit{low }level of articulation. While we did not explicitly support voice-to-text, as we were primarily interested in written reflection, screen recordings captured user audio by default, allowing for future transcription and voice-to-text alignment.
    \item \textbf{Level of review}: SavePoint supported \textit{low} and \textit{medium} levels of review. After each session, players were automatically presented with a gallery window (see Figure \ref{flow}) displaying session documents (low level). We provided users the option to delete individual annotations from a session, delete entire sessions, or provide unstructured post-session reflections (medium level). The app did not explicitly prompt high levels of review; rather, we gave participants structured reflection questions to answer outside of the app in moderated social discussions. 
\end{itemize}

In terms of SAVEstate's technical considerations, SavePoint used the following:

\begin{itemize}
    \item \textbf{Naturalistic collection}: To generate gameplay streams, SavePoint utilizes the Open Broadcast Studio (OBS) API. We chose OBS due to its pre-existing integrations into the gaming ecosystem. 
    \item \textbf{Processing the anchors}: The app generated three types of data: 1) annotations (which contained the Annotation list as a JSON file), 2) session information (which contained the Session start [timestamp], Game name [string], and Date of play [timestamp] as a single JSON file), and 3) videos (.mkv files). 
    \item \textbf{Data collection and post-processing}: We used Amazon Web Services Simple Storage Solution (AWS S3) as our solution for centralized data collection. 
\end{itemize}

% \subsection{User Flow}
% Below, we describe how users interact with SavePoint in a typical app session (also displayed in Figure \ref{flow}). 
% \begin{enumerate}
%     \item If user is opening SavePoint for the first time, they are prompted to enter a username. This saves a configuration file within the user's system files. If they have used the app before, they are brought straight to the landing page. 
%     \item From the landing page, the user chooses to "Start New Session". They are prompted for what game they are playing. Once submitted, the main process sends a StartRecord call to OBS and generates a timestamp. A metadata JSON file containing the game name, username, and starting timestamp are written to S3. An empty JSON file for annotations is separately generated in S3. 
%     \item The user uses the hotkeys for text (Control/Command + Shift + N) or emote (Control/Command + [1-6]) based inputs to create annotations over the course of their session. When an annotation is made, it is timestamped and saved to the annotation JSON file in S3.
%     \item When the user wishes to finish the session, they press the quit hotkey (Control/Command + Shift + Q). The main process sends a StopRecord call to OBS. To provide user feedback while the video is being written to the disk and then uploaded to the cloud, the app monitors state and shows the user a loading window. 
%     \item Upon completion of the upload, the gallery window appears and users can review their session. 
%     \item Users quit the app from the landing page or closing out of the gallery window. Either action initiates the main process to terminate all child processes to ensure a clean exit. 
% \end{enumerate}

\section{Case Studies}

\subsection{Overview of Case Studies} The following case studies investigated SAVEstate use while varying the levels of the method's three main considerations: scaffolding, articulation, and review (see Table \ref{table:case-study-configurations}). Choosing different levels for each consideration allowed us to test the method's adaptability across variable circumstances. We chose the cases based on what we believed were SAVEstate's primary affordances, namely (1) prolonged engagement with a single game, (2) in-situ reflection in natural environments, and (3) scaffolding dialogical reflection. We also wished to investigate how SAVEstate functioned in a range of individual and social settings; namely, individual reflection-in/on-action (Case 2), individual reflection-in-action and social reflection-on-action (Case 1), and social reflection-in/on-action (Case 3). Finally, Cases 1 and 2 were conducted remotely both for participant convenience and to test SavePoint's robustness as a scalable data collection tool for researchers, while Case 3 was conducted in-person due to the social nature of the research question. We deployed Case 3 through foundry10's\footnote{A philanthropic research foundation which creates youth programming for games and education.} youth cohort program, for which the first and fifth author are both instructors. 

\begin{table}[h]
\centering
\caption{SAVEstate design configurations and settings across three case studies.}
\label{table:case-study-configurations}
\footnotesize
\begin{tabularx}{\textwidth}{@{} l p{1.2cm} p{3.1cm} p{3.1cm} p{1.5cm} p{1.5cm} p{1.5cm} @{}}
\toprule
\textbf{Case ID} & \textbf{Setting} & \textbf{Reflection-in-action} & \textbf{Reflection-on-action} & \textbf{Scaffolding} & \textbf{Articulation} & \textbf{Review} \\ \midrule

\textbf{1} & Remote & Individual & Social & Medium & High only & High \\ \addlinespace \midrule

\textbf{2} & Remote & Individual & Individual & Low & Low, High & Low \\ \addlinespace \midrule

\textbf{3} & In-person & Social & Social & Medium & All & Medium \\ \bottomrule
\end{tabularx}
\end{table}

All the case studies were approved by institutional (Cases 1 and 2; STUDY00023769) or organizational (Case 3) review boards. Participants (Case 2) or their parents (Case 1) completed the approved consent form prior to the beginning of the study. For Case 3, data collection for the youth cohort program was pre-approved through an organizational IRB (\#2025-017).

\subsubsection{Case 1: Game Book Club} The first context in which we tested SAVEstate was a game book club with high school students. The underlying research question was: Can we use SAVEstate to explicitly connect in-game reflections to externally moderated reflection-on-action? We recruited participants through convenience sampling, sending fliers through university Slack channels and mailing lists for youth interested in game related studies. We conducted a month long study in which 8 high schoolers played \textit{Tacoma} \citegame{tacoma}, a story-driven sci-fi game set aboard an abandoned lunar space station, where players must discover the mystery behind its evacuated crew. In the reflection-in-action phase, we provided medium researcher-guided scaffolding, instructing students to annotate with an eye toward solving the game's underlying mystery. The reflection-on-action phase required high levels of review to answer weekly discussion questions in a Discord sever, primarily concerning the game's main themes of AI ethics and corporate malpractice. We purchased copies of the game for all participants and additionally compensated them with gift cards worth \$30 USD. 

\subsubsection{Case 2: Playing Critically} Next, we tested SAVEstate in a two-week naturalistic study. The underlying research question was: Can SAVEstate help us observe micro-loops of meaning-making in players' everyday gaming habits without the intervention of a researcher or social group? We recruited 17 adult participants using convenience sampling and asked them to integrate SavePoint into their regular gaming routines for a 2 week period. Participants were asked to complete at least three gaming sessions while using the tool. In the reflection-in-action phase, we provided low scaffolding, instructing participants to annotate in a way that felt natural to them. The reflection-on-action phase enabled medium levels of review. We also sought to investigate whether users believed that SavePoint supported their reflective capabilities and was generally non-intrusive to their natural play. To do so, we asked users to evaluate the tool on the Reflection, Rumination, and Thought Scale \cite{Loerakker_Niess_Wozniak_2024} and the System Usability Scale \cite{lewis2018system}, as well as qualitatively detail if they would use the tool in their everyday practices and why. Participants were each compensated gift cards worth \$20 USD. 

\subsubsection{Case 3: Collective Debriefing in Educational Games}
Finally, we tested SAVEstate in a classroom setting in which students collaboratively played an educational game. The underlying research question was: Can we use SAVEstate to support social dialogical in-game reflection in a classroom setting? We conducted an hour long session with 10 students aged 11-14 as a part of foundry10's youth cohort program. Students used SavePoint while playing \textit{Breaking Harmony Square} \cite{roozenbeek2020breaking} (an educational game about misinformation and fake news in which players primarily interface with humorous social media posts) in four small groups. In the reflection-in-action phase, we provided high researcher-guided scaffolding, explicitly instructing students to annotate when an aspect of the game reminded them of real life. In the reflection-on-action phase, we had students go around the classroom and view other groups' reflections before concluding the session with a full class discussion. 

\subsection{Case Study Analysis Methods}
We collected a total of 1631 contextual anchors over 2440.5 minutes of gameplay video across all three case studies (see Table \ref{table:case-studies}). We also collected written exit interviews for Cases 1 and 2, Discord discussion posts from Case 1, and audio transcriptions from the debrief sessions in Case 3.

We used \citet{marshall1997annotation}'s functions of annotations as a guiding deductive framework for examining the contextual anchors because of its salience to motivations for designing SAVEstate; namely, it outlines ``the value [of personal annotations] to the annotators and to later readers [e.g. the researchers], their functions, and the implications [of] existing practice''. To classify annotations under each of \citet{marshall1997annotation}'s functions, the first author placed 10\% of the data (163 notes in 75 groupings, drawn from all 3 cases) into a Figma board. The first and the fifth author then collaboratively affinity diagrammed \cite{scupin1997kj} each grouping of notes into the best aligned corresponding function. The second author additionally provided notes on how to adapt \citet{marshall1997annotation}'s definitions for our context and observed which functions corresponded to which study cases. The outcomes of this process are discussed in Section \ref{diverse}. 

To generate themes surrounding the affordances and tensions of SAVEstate (Sections \ref{afford} \& \ref{tensions}), the first and fifth author used thematic analysis \cite{clarke2017thematic}. We returned to the aforementioned subset of data (with which we were already familiarized) and performed two rounds of review: one to develop preliminary themes, and a second to come to consensus on a working set of themes. Afterwards, the first author re-reviewed the entire dataset to refine and develop final definitions for themes. 

In extracting representative quotes and descriptions for each of \citet{marshall1997annotation}'s functions, as well as our themes, we used the "vivid exhibits" lens \cite{cscwreport} utilized in ethnography. We found this appropriate as our data was similar to that generated in ethnographic studies, namely notes, audio, and video recordings from naturalistic settings \cite{lecompte2012analysis}. Themes that were well represented with numerous examples across studies were included here, illustrated with exhibits from the larger collection.

In the following sections, we refer to participants by their participant ID and the case they participated in. For example, participant with PID \#8 in Case 2 would be referred to as P8-C2.

\begin{figure}
  \centering
  \includegraphics[width=\linewidth]{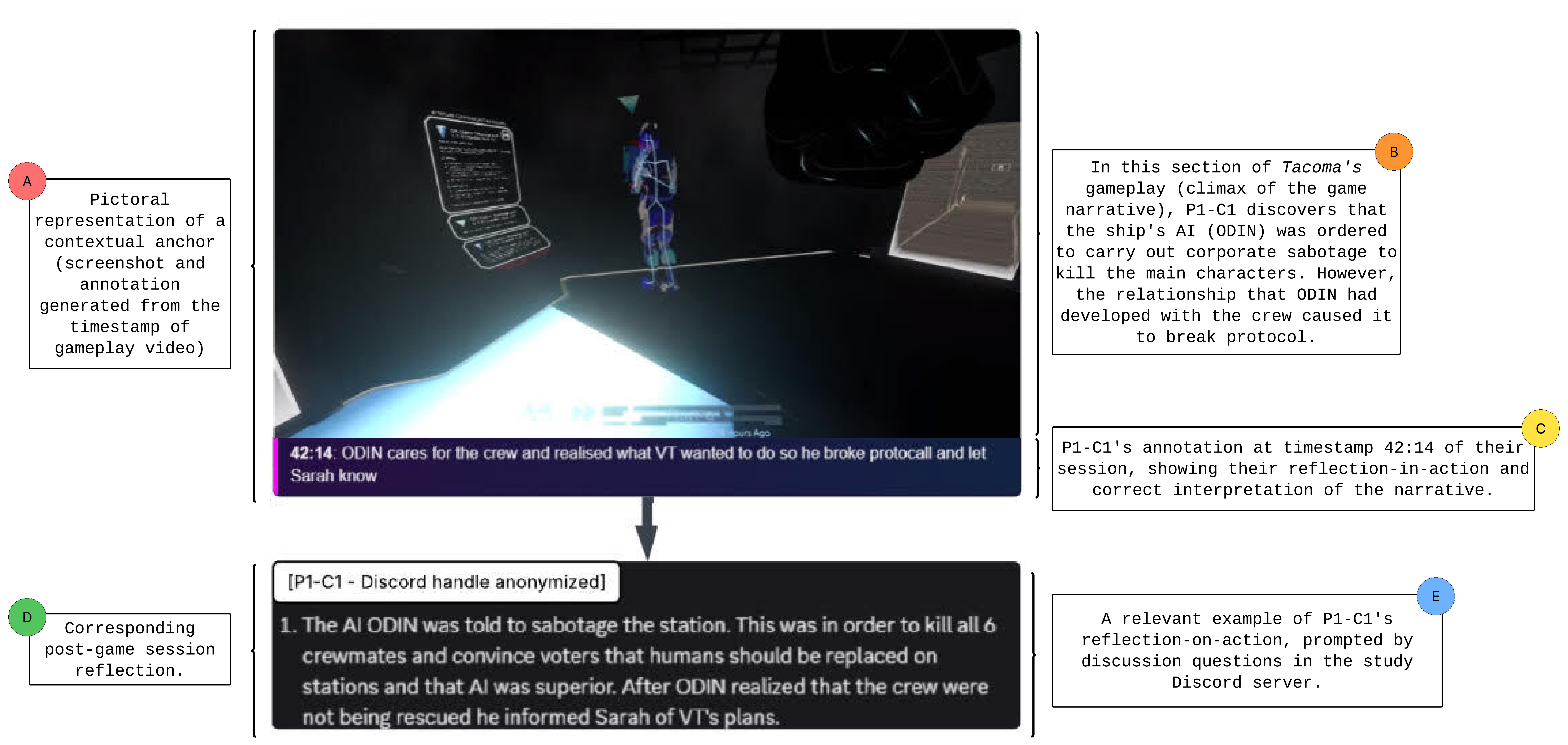}
  \Description{Explanation of contextual anchor}
  \caption{A pictoral representation of a contextual anchor from P1-C1's gameplay of \textit{Tacoma}. The screenshot is taken from the player's gameplay stream at the timestamp of the in-situ annotation (Note C). The researchers then connected the contextual anchor to P1-C1's post-game reflection, which is a snippet from their discussion question responses in the study Discord server. We use this pictoral format to display contextual anchors and related post-game reflections throughout the Findings section.}
  \label{explanation}
\end{figure}

\section{Findings}

\begin{table}[h]
\centering
\caption{Overview of case studies showing the study goals, participant count, age ranges, study duration, notes taken (emote/text), average words per annotation, and total length of video generated.}
\label{table:case-studies}
\small % Slightly smaller font to help with fit
\setlength{\tabcolsep}{4pt} % Tighten horizontal space
\begin{tabularx}{\textwidth}{@{} l X c c c c c c @{}}
\toprule
\textbf{ID} & \textbf{Goal} & \textbf{N} & \textbf{Age} & \textbf{Dur.} & \textbf{Notes (E/T)} & \textbf{Avg. Words} & \textbf{Vid (m)} \\ \midrule
\textbf{1} & Connect in-game reflection to social discussion & 8 & 14--18 & 4 wks & 347 (18/329) & 15.4 & 512.6 \\ \addlinespace
\textbf{2} & Capture micro-loops of single-player meaning-making & 17 & 18--40 & 2 wks & 1193 (512/681) & 11.6 & 1789.5 \\ \addlinespace
\textbf{3} & Investigate social meaning-making in classroom setting & 10 & 11--14 & 1 hr & 91 (55/36) & 9.5 & 138.4 \\ \bottomrule
\end{tabularx}
\end{table}

\subsection{Evidence of Diverse Documentary Practices} \label{diverse}

\begin{table}[ht]
\centering
\caption{Cross-mapping of \citet{marshall1997annotation}'s forms and functions of annotations and how we adapted them for SAVEstate.}
\label{table:combined-annotations}
\footnotesize
\renewcommand{\arraystretch}{1.5}
% Increased third column to p{3.2cm} for a cleaner look
\begin{tabularx}{\textwidth}{@{} p{2.8cm} X p{3.2cm} X c @{}}
\toprule
\textbf{Function (Original)} & \textbf{Form (Original)} & \textbf{Function (SAVEstate)} & \textbf{Form (SAVEstate)} & \textbf{Cases} \\ \midrule

\rowcolor{tablepurple}
Procedural signaling for future attention & Underlining or highlighting higher level structure (like section headings); telegraphic marginal symbols like asterisks; crossouts. & Procedural signals for marking intended action or strategic navigation & Outlining high level strategy, e.g. "gonna try and beat boss rush and Hush with Lazarus" (P5-C2); typographically emphasizing thoughts to self, e.g. "i guess i have to Absolutely Not Jump there" (P2-C2) & 1,2 \\

Placemarking and aiding memory. & Short highlightings; circled words or phrases; other within-text markings; marginal markings like asterisks. & Placemarkings for information to return to later & Short summaries of game sections, e.g. "So to recap: EV and Clive are in cryo to save oxygen, Andrew is soon to follow, Sareh is probably going to need to sit tight, I assume Nat is going into cryo as well, and Bert is going to get help" (P8-C1); metaphorically pinning parts of game to return to, e.g. "Let's put a pin in the fleshy door for later lol" (P1-C2) & 1,2 \\

\rowcolor{tablepurple}
Interpretation & Short notes in the margins; longer notes in other textual interstices; words or phrases between lines of text. & Interpretive activity signaling endo/exo-game meaning making & Eureka moments of understanding (e.g. Figure \ref{meaning_1}); references to previous parts of the game or past playthroughs, e.g. "I've never noticed it this early in the game, but the references to 'wanderers' are really well-ingrained throughout the whole plot." (P15-C2) & 1,2,3 \\

Problem-working & Appropriate notation in margins or near figures or equations. & Problem-working & Using note area as thinking space for puzzles, e.g. Figure \ref{p1_puzzle}. & 1,2 \\

\rowcolor{tablepurple}
Visible traces of attention & Extended highlighting or underlining. & Traces of affect & Emote reactions (e.g. Figure \ref{case3_funny}); keyboard smashes or all-caps annotations (e.g. Figure \ref{meaning_2}) & 2,3 \\

Incidental reflection of the material circumstances of reading & Notes, doodlings, drawings, and other such markings unrelated to the materials themselves. & Incidental reflections of the material circumstances of playing & Information unrelated to the game and reflective of physical status, e.g. hunger or location & 2 \\ \bottomrule
\end{tabularx}
\end{table}
\begin{figure}
  \centering
  \includegraphics[width=0.65\linewidth]{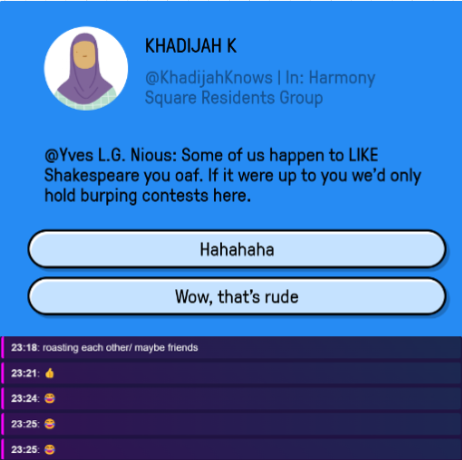}
  \Description{P1-C3 and P2-C3 annotated that this in-game social media post reminded them of "roasting each other/maybe friends". They also annotated several laugh emotes. In the audio from the corresponding section of the generated gameplay stream, they are audibly giggling.}
  \caption{P1-C3 and P2-C3 annotated that this in-game social media post reminded them of "roasting each other/maybe friends". They also annotated several laugh emotes. In the audio from the corresponding section of the generated gameplay stream, they are audibly giggling.}
  \label{case3_funny}
\end{figure}

In adapting \citet{marshall1997annotation}'s functions to a game context (Table \ref{table:combined-annotations}), we retained the core structure of each function while reorienting them around the specific affordances and constraints of gameplay. Functions that were originally tied to the physical materiality of text, such as underlining, highlighting, and marginal symbols, were reinterpreted as temporal and strategic markers within an interactive medium. In this vein, we reinterpreted Function 5 (visible traces of attention) as traces of affect as afforded by the emote reactions in SavePoint as we believed this was a more useful function to the annotation of games specifically. Although we did not deeply investigate the use of voice notes in the deployed SavePoint prototype, this would add additional forms. For example, a player creating a voice note specifically to verbalize frustration would function as a trace of affect. 

\begin{figure}
  \centering
  \includegraphics[width=\linewidth]{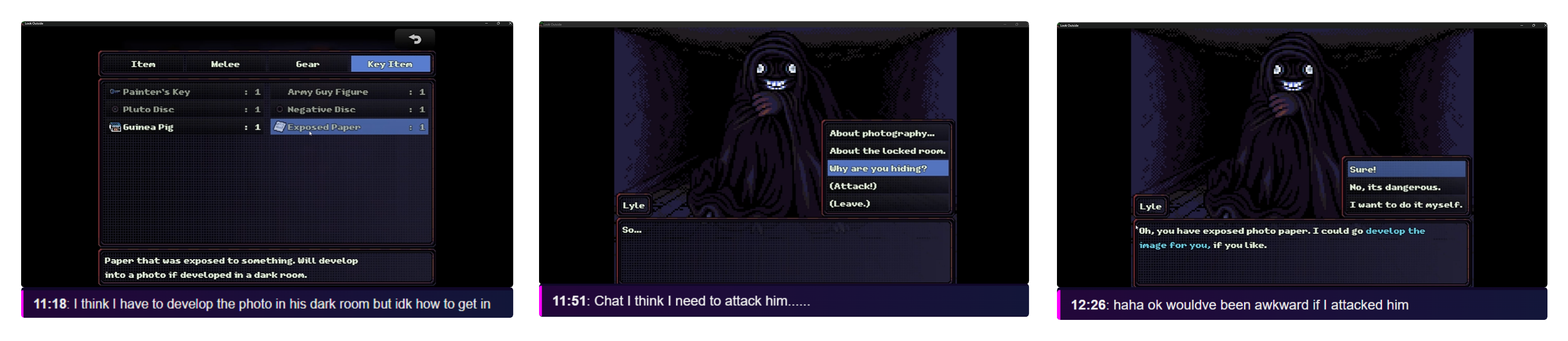}
  \Description{A sequence of contextual anchors showing P1-C2 determining how to accomplish an in-game task.}
  \caption{A short sequence of P1-C2 problem-solving a task in \textit{Look Outside} \citegame{lookoutside}, a survival horror game. In the first contextual anchor, P1-C2 is examining their item inventory and annotates that they "have to develop the photo in [the character's] dark room" but they do not know how to enter. The photo is a necessary object for progressing in the game. In the second contextual anchor, after entering the character's room, they wonder if they have to attack him after being presented with the option. However, the proceeding section of P1-C2's gameplay shows that the character (despite appearances) is quite friendly and will help them develop the photo. P1-C2 annotates relief in the third contextual anchor that they did not attack the character.}
  \label{p1_puzzle}
\end{figure}

\begin{enumerate}
    \item \textbf{Annotations as procedural signals}: These annotations were common in Cases 1 and 2 and primarily took the form of participants giving themselves instructions for what actions to take in particular moments. Often, these annotations served as useful signals for the researchers as to what participants were doing in sections of the games, which was especially useful for games that the research team was unfamiliar with. P5-C2, for example, started every run of their \textit{The Binding of Isaac: Repentance} \citegame{bindingofisaacrepenetance} sessions with a statement about which bosses they would try to beat with which character. 
    \item \textbf{Annotations as placemarkings and aids to memory}: These annotations primarily manifested as the recording of pieces of information that participants wished to recall later. For example, \textit{Tacoma} required players to find a number of passcodes to progress, and participants used the annotations to preserve them for memory. Similarly, all participants annotated extensively on the characters in \textit{Tacoma}, as the characters had different jobs and relationships that were difficult to track. In Case 2, participants who played puzzle or exploration-heavy games used the annotations as reminders. Notably, SavePoint was designed such that contextual anchors could only be reviewed at the end of a play session so placemarkings could not be reaccessed during play. Several participants in Cases 1 and 2 requested "a sidebar" (P14-C2) or "self-chat" (P17-C2) to "view annotations before [they] finish the session" (P1-C1). 
    \item \textbf{Annotations as record of interpretive activity}: These annotations were universal across all case studies, and as they were the main targets of our method development, we discuss several examples in depth in Section \ref{meaningmaking}.
    \item \textbf{Annotations as in-situ locations for problem-working}: These annotations primarily surfaced in Cases 1 and 2 through solving in-game puzzles or mysteries. For example, Figure \ref{p1_puzzle} shows P1-C2 using the annotations as thinking space for accomplishing a task (develop a photo). The contextual anchors also demonstrated breakdowns in participants' problem solving. P8-C1, in their attempt to deduce a story element, spent around 10 minutes attempting to find the requisite password. They began by looking around the room, annotating, "There's gotta be more to this one, right?" After three more minutes of searching, they located the correct hint and annotated, "WAIT! It was in front of me the whole time." However, they misunderstood the hint, and eventually annotated an incorrect reasoning behind the story element due to never finding the passphrase.
    \item \textbf{Annotations as traces of affect}: These annotations denoted obvious emotions that underscored particular in-game moments. These included keyboard smashes or all-caps messages (e.g. Figure \ref{meaning_2}), or liberal emote usage (e.g. Figure \ref{case3_funny}). Sadness represented 25\% of emote use across all studies. Five participants in C2 preferred low articulation to high; P16-C2 annotated almost entirely in terms of their affect, with emotes making up 98.8\% of their total annotations. 
    \item \textbf{Annotations as incidental reflections}: These annotations only surfaced in Case 2. Almost all the instances we observed were of participants reporting that they were hungry (P2, P12, P15) or were leaving their keyboard ('going AFK') to deal with something in their surroundings (P5, P15). Although these types of annotations were relatively rare, they likely could only surface during informal play where participants felt comfortable to record information unrelated to the game.
\end{enumerate}

\subsection{Affordances of SAVEstate} \label{afford}

\subsubsection{Observing instances of meaning-making} \label{meaningmaking}
\begin{figure}
  \centering
  \includegraphics[width=0.7\linewidth]{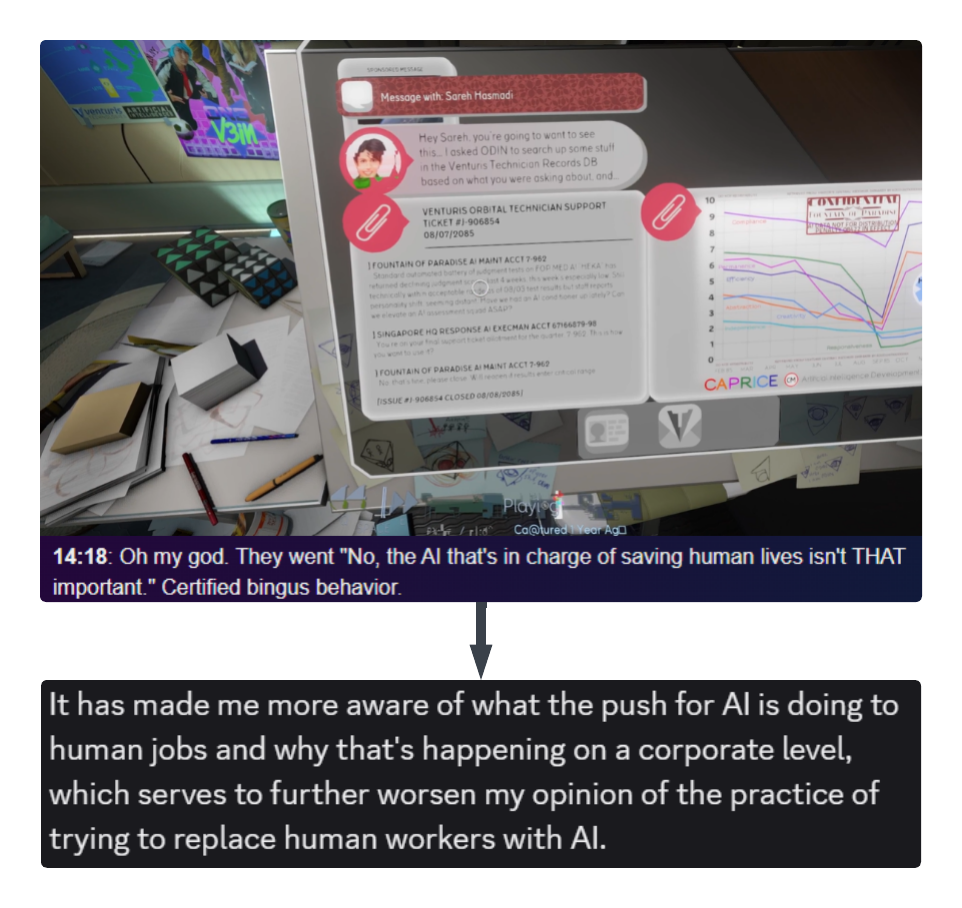}
  \Description{A contextual anchor (top image) from P8-C1 showing how they interpreted a series of in-game messages and graphs to conclude an important plot point in \textit{Tacoma}. The graph shows that despite an obvious decline in a medical AI's performance, the antagonist company ignored the warning signs due to performance not being in critical range. This led to one of the main characters, a doctor, being unjustly blamed for a patient death. In their proceeding reflection-on-action in the Discord discussion (bottom image), P8-C1 stated that this point in the game "further worsen[ed] [their] opinion of the practice of trying to replace human workers with AI".}
  \caption{A contextual anchor (top image) from P8-C1 showing how they interpreted a series of in-game messages and graphs to conclude an important plot point in \textit{Tacoma}. The graph shows that despite an obvious decline in a medical AI's performance, the antagonist company ignored the warning signs due to performance not being in critical range. This led to one of the main characters, a doctor, being unjustly blamed for a patient death. In their proceeding reflection-on-action in the Discord discussion (bottom image), P8-C1 stated that this point in the game "further worsen[ed] [their] opinion of the practice of trying to replace human workers with AI".}
  \label{meaning_1}
\end{figure}

\begin{figure}
  \centering
  \includegraphics[width=\linewidth]{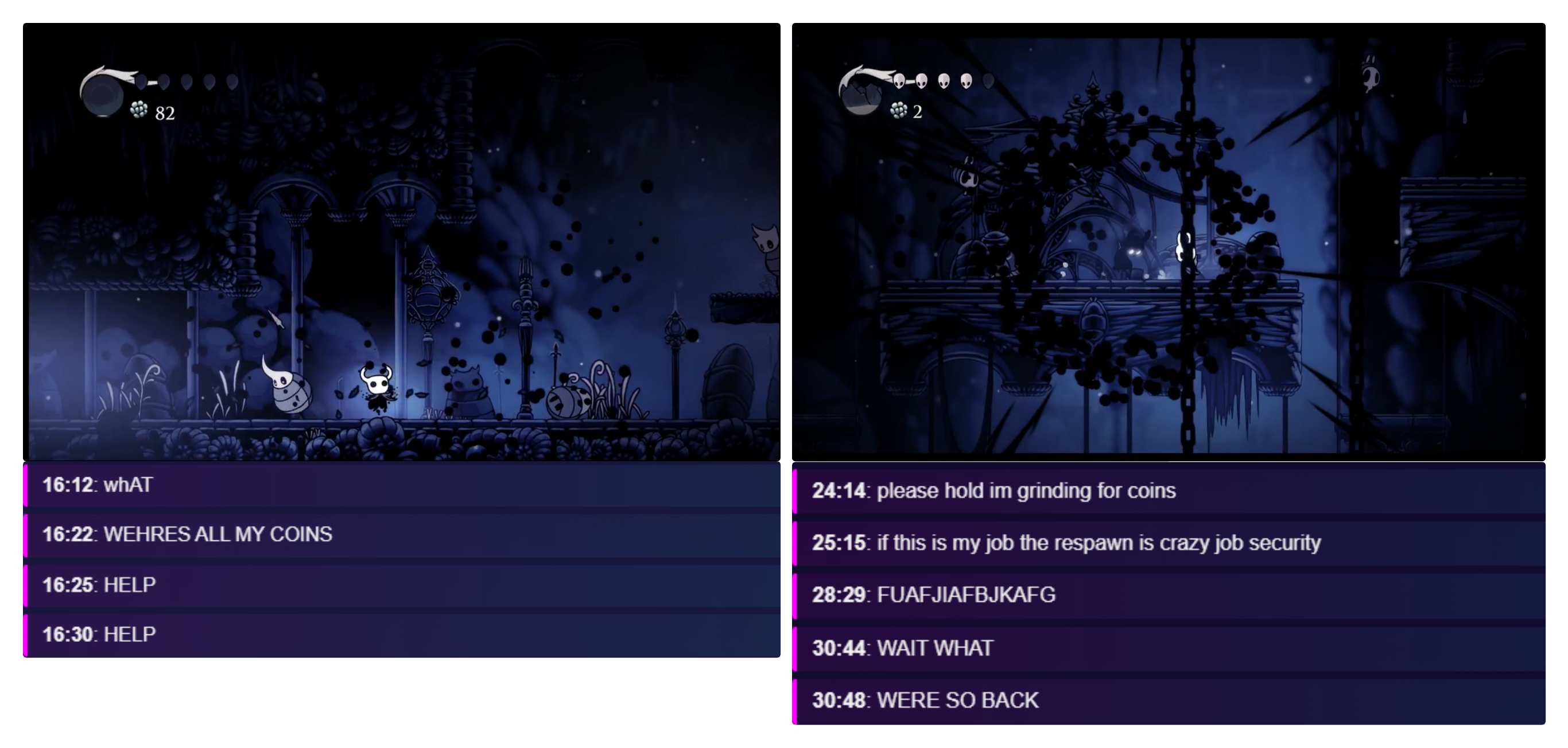}
  \Description{P17-C2}
  \caption{P17-C2 piecing together that they can reclaim resources by returning to the location they last died at, shown through contextual anchors from two sessions playing \textit{Hollow Knight} \citegame{hollowknight}.}
  \label{meaning_2}
\end{figure}

Across all three case studies, we routinely found that the contextual anchors provided a lens into how participants interpreted aspects of the game and how they chose to reflect upon it. In Cases 1 and 2, we primarily observed how participants made \textbf{endo-game reflection-in-action} \cite{Whitby_Iacovides_Deterding_2023}. In the example shown in Figure \ref{meaning_1}, for the minute preceding the annotation, P8-C1 interpreted the in-game messages and graph to determine that the evil corporation in \textit{Tacoma} purposefully overlooked the degradation of an AI model for health decisions, which led to negative outcomes for a main character. Figure \ref{meaning_2} shows a series of annotations from P17-C2 across two sessions. P17-C2 initially appeared unaware of \textit{Hollow Knight}'s Souls-like mechanics, namely that upon dying, the player loses all their coins and must find the location in which they last died to reclaim them. While P17-C2 did not realize this in their first session (left side, Figure \ref{meaning_2}), their contextual anchors in the second session (right side, Figure \ref{meaning_2}) tell us how they eventually came to understand the mechanic. They initially annotated that they are grinding for coins (24:14), presumably to purchase items from the shop that they discovered earlier in the session. However, they died (28:29), and returned to the area to restart their objective of collecting coins. Here, they discovered that reclaiming their character's soul returns their money (30:44-30:48).   

\begin{figure}
  \centering
  \includegraphics[width=0.7\linewidth]{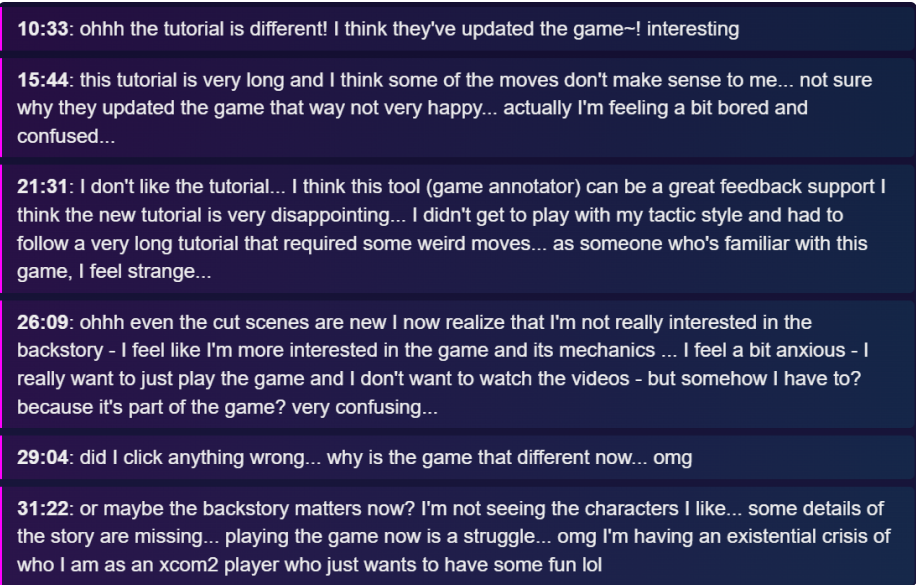}
  \Description{P7-C2}
  \caption{A series of annotations from P7-C2 realizing, upon replay of XCOM 2 \citegame{xcom2}, that the game had changed significantly since the last time they had played it.}
  \label{p7}
\end{figure}

In Case 2, where participants chose the games they played, we saw participants make connections back to previous playthroughs, shedding insight on \textbf{replay}: ``how [previous playthroughs] feed back into play and affect subsequent reflection-in-action'' \cite{Mekler_Iacovides_Bopp_2018}. P15, for example, annotated that they have "replayed [A Dark Room] \citegame{adarkroom} maybe 20-30 times...but this game's plot is genuinely really good and it still makes [them] feel insanely happy whenever [they] finish it again". They connected back to memories from previous playthroughs, e.g., "Love the reveal of the 'wanderer fleet', when I first played this the realization was actually insane", and annotated on new aspects of the game they noticed since their last playthrough, e.g., "I've never noticed it this early in the game, but the references to 'wanderers' are really well-ingrained throughout the whole plot." In a combat sequence where P3 was fighting crab-like monsters in \textit{World of Warcraft} \citegame{wow} after a period spent away from the game, they annotated: "remembering now that I used arachnophobia mode to get rid of the spiders...I wish more games had that option". The researchers, upon investigating this, discovered that arachnophobia mode in \textit{WoW} turns spider enemies into the crab-like monsters we saw in the corresponding section of the stream. Several of P7's annotations documented a slow realization that \textit{XCOM 2} \citegame{xcom2} had changed significantly since their original playthrough, and this caused them confusion and anxiety (Figure \ref{p7}). 

Finally, we primarily saw instances of \textbf{exo-game reflection-in-action} (see Section \ref{sense}) in Case 3, likely due to the game's realism and scaffolding through the activity instructions. These instances of interpretation were bolstered by participant discussion. Notably, participants did not always discuss exactly what they annotated; the activity scaffolding appeared to serve as a forcing function for stabilizing their reflections. For example, early on in Group 4's session, they read a fake post which stated, "Certain vaccines are loaded with dangerous chemicals and toxins." After marking it as false, they moved onto the next post and began to read it. Then P9 cut in saying, "Oh wait, we're supposed to take a note when something reminds us of real life," and scrolled back to the vaccine post to make the annotation, "vaccines having chemicals and toxins relates to social media posts by our government about vaccines". We did not observe notable instances of exo-game reflection in the contextual anchors from Cases 1 or 2. 

\subsubsection{Connecting anchors to reflection-on-action}
\begin{figure}
  \centering
  \includegraphics[width=0.7\linewidth]{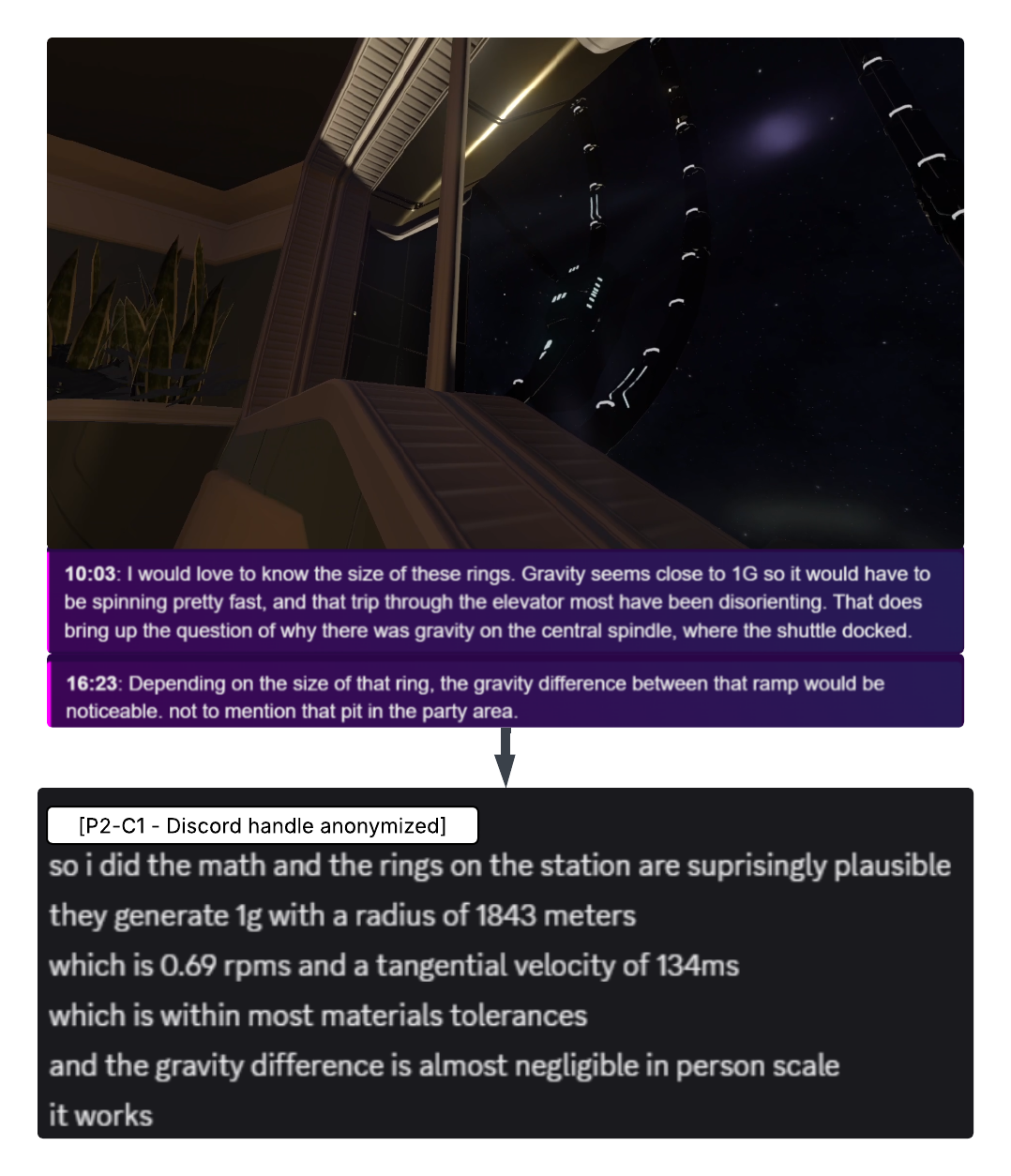}
  \Description{Connecting P2-C1's contextual anchors to later discussion in the Discord server.}
  \caption{Connecting P2-C1's contextual anchors to interest-driven discussion in the Discord server. Upon observing the gravity rings on the \textit{Tacoma} spaceship, P2-C1 wondered about their feasibility and started a discussion about it in the Discord.}
  \label{p2c1}
\end{figure}

As Cases 1 and 3 contained moderated post-game activities, using SAVEstate enabled us to connect contextual anchors to later discussions. In Case 1, these were often straightforward translations from contextual anchors to answering weekly discussion questions in the Discord server, as shown in Figure \ref{meaning_1}. Annotations also provided insight into interest-driven discussions that participants engaged in within the Discord outside of their assigned questions. For example, as shown in Figure \ref{p2c1}, P2-C1's curiosity about the artificial gravity rings on the \textit{Tacoma} spaceship manifested in them starting a thread with other participants on their plausibility. 

In Case 3, the contextual anchors shifted from private reflections to objects of social discussion when participant groups were instructed to review a different group's session. Participants primarily used the anchors to judge the strategic choices of their peers, especially with regards to how many points the other group got. For example, when P9 and P10 from Group 4 reviewed Group 1's annotations, P10 initially commented, "We got a better high score", to which P9 replied, "they were too nice...sometimes [the game] gives you two bad options and you can't pick a good one. So I think they tried to be nice, but the game kept forcing them to try and get likes". Group 3, when reviewing Group 2's session, noticed that Group 2 had significantly less points than them. P6 from Group 3 laughed as they said, "I think they were like, what do we do!" P7 said, "They have way less likes than us," to which P6 immediately responded, "Yeah, they suck!" In the other room, P4 from Group 2, when reviewing Group 3's session, said with disdain, "They didn't care [what options they were choosing]! They barely read anything!" Participants also commented on the quality of each other's annotations, independent of their gameplay. P1 from Group 1, when reviewing Group 4's session, stated that, "They did a lot better on the notes than we did". Indeed, Group 4 had followed the instructions carefully and taken thorough annotations. Conversely, P4 from Group 2 negatively judged Group 3's annotations, saying, "I feel like this is not very deep commentary, no hate. Like, ours was deep."  

%emmett ginny - group 1
%millie lita girl - group 2
%pierce jack and marina? - group 3
%pika and other kid - group 4

% \begin{figure}
%   \centering
%   \includegraphics[width=0.7\linewidth]{p1c1.png}
%   \Description{Connecting P1-C1's contextual anchors to later discussion in the Discord server.}
%   \caption{Connecting P1-C1's contextual anchors to their answers to assigned weekly discussion questions about the game.}
%   \label{p1c1}
% \end{figure}

\subsubsection{Viewing sensemaking synchronously} \label{sense} In Cases 1 and 3, using SAVEstate helped us quickly identify common themes across participants' reflection-in-action. For example, in Case 3, where participants were specifically instructed to annotate when something in the game reminded them of real life, all four groups made connections back to modern day American politics. The premise of the game is introduced as a location in which "three political parties are bickering constantly" \cite{roozenbeek2020breaking}, to which Group 4 annotated, "This sounds like our government today with many different parties who don't agree with each other". As the player character is tasked with spreading misinformation, Group 2 commented, "fake news being spread about elections reminds me of real life". When the player character's actions start to cause turmoil in the game world, Group 3 annotated, "when you turn an issue into a heated debate, it becomes a fight, which causes panic, which cause[s] [the] mafia [to] overtake [the] government". At the end of the game, where the player character's actions finally escalate to causing an outbreak of civic unrest, Group 1 annotated, "The news [in January 2026] is always talking about the National Guard". 

SAVEstate also helped us recognize which moments in the game prompted comments or reflections for a majority of players. For example, most participants in Case 1 left annotations at the climax of \textit{Tacoma}, where it is revealed that the incident the main character has been investigating was no accident, but rather corporate sabotage. P6 annotated their moment of discovery: "AH HA, ODIN was ordered to cause the accident". P2 ("I knew it") and P4 ("oh, so we're the clean up crew...") had more cynical reactions, while P7 successfully connected the event back to the broader context of the game: "WOW, so this was for sure staged to turn these stations into ones using only AI". In Case 3, as players scrolled through in-game social media arguments, three groups made very similar comments on what the contents reminded them of: "Sounds like two political rivals arguing" (Group 4), "Sounds like a Karen" (Group 2), and "Sounds like Trump or Musk" (Group 1). 

\subsection{Tensions of Using SAVEstate} \label{tensions}

\subsubsection{Tensions between annotating naturally and feeling observed} \label{tensions1}
From a researcher perspective, allowing participants to annotate in a way that felt completely natural to them came with tradeoffs. In some cases, participant documentation of a game that they had replayed many times was high context for them but low context for the researchers, which made it difficult to parse the contextual anchors without knowledge of the game itself. For example, P5-C2 and P11-C2 both documented \textit{The Binding of Isaac: Repentance} \citegame{bindingofisaacrepenetance}, a roguelite with a highly complex mechanics system that the research team was unfamiliar with. P11-C2 made references to rerolling on certain items that bestowed power-ups to their character, but as the item name was not provided in the annotation or the recording, it was difficult to determine its meaning. Similarly, P5-C2 would start each session with an annotation about what character they chose and what boss they were attempting to beat, but without deep knowledge of the game, it was difficult to determine what this meant to the player.

Conversely, we observed that several participants in Case 2 annotated as if they were communicating with an external observer. This typically occurred as providing additional context to the researchers for what was occurring in the game. For example, when using a particular mechanic in \textit{A Dark Room} \citegame{adarkroom}, P15 annotated: "For context I've been playing this game on and off since middle school and still only recently discovered the Fabricator. I don't get bored playing this because I just keep finding new lore/mechanics". P2, after reaching a passage between two areas in \textit{Rain World: The Watcher} \citegame{rainworld}, first annotated, "KARMA GATEEEEEE", and then explained to the researcher in the next annotation, "so in rain world these are the only gates between areas. to get through them you have to stay alive for a certain number of consecutive cycles (days) without getting Eaten By Lizards". Some participants even explicitly used language drawn from the broader game streaming ecosystem, such as P13: "What's up what's up. Today we will be trying out a new game that I have not played before called Realm of the Mad God \citegame{realmofthemadgod}".
%While playing \textit{Mosaic of the Pharaohs}, a game visually and mechanically similar to \textit{Minesweeper}, P12 annotated on how they believed the researchers would perceive their data: "I keep thinking this will be the most boring game to watch because there's no ppl in it, it's just me clicking buttons". 

Prior work examining other in-situ methods has shown that participant knowledge of data collection can create a Heisenberg Effect; the process of self-reporting experiences can change what participants do and how they feel about it \cite{czerwinski2004diary}. Indeed, all participants were informed that the research team would view their documents as part of our analysis, and as a result, some participants in C2 noted that the process of documenting made them ``put on a performance'' (P12-C2) more than they would have otherwise. P17 in C2 noted feeling "like [they] were interacting with a sort of pseudo-audience", and P1 in C2 believed that they were "a bit more aware of an external observer than myself as the observer...which distracted from more self-focused reflection". We did not receive similar feedback in C1 or C3, likely because the games were played in informal educational spaces where the expectation of observation was more justified.

\subsubsection{Tensions between documenting and playing} \label{tension2} Some participants spoke positively about reflecting-in-action, commenting on "trying to be more aware" (P3-C2), having to "stop and think about the game more" (P4-C1), and that "having to annotate a moment...lets [them] slow down [and] think about how [they] are spending [their] time" (P13-C2). Indeed, for some participants, documenting helped "organize [their] thoughts and see how [their] thinking [changed] as the game progressed" (P1-C1) and "made the session[s] feel more intentional", as if they were spending "more direct time...doing a specific activity" (P11-C2). The documents served as "journal[s] in which [players] could unpack what [they] were thinking/feeling" (P15-C2). Participants benefited from the reflection-on-action phase of the method as well, stating that "it was nice to see how long [they] had been playing for and have that knowledge" (P10-C2). Interacting with their past sessions helped them "feel less like [their] time was going into a void" (P6-C2) and "made each moment more important" (P13-C2). These findings are supported by results from the R2T2 scale. Participants in Case 2 ranked SavePoint highly for supporting reflection, believed it did not promote negative rumination, and found it adequate for supporting self-focused thinking (see Table \ref{r2t2}). These results speak to the feasibility and acceptability of using tools like SavePoint for supporting SAVEstate. 

\begin{table}[ht]
\centering
\small
\renewcommand{\arraystretch}{1.4} % Comfortable row spacing
\begin{tabular}{l p{8cm} c}
\toprule
\textbf{Question ID} & \textbf{Question} & \textbf{Mean (SD)} \\ \midrule
\rowcolor{tablepurple}
REF1 & Using this technology made me conscious of my behaviors. & 4 (0.61) \\
REF2 & This technology helps me to be able to reflect more easily on my actions. & 3.76 (0.83) \\
\rowcolor{tablepurple}
REF3 & This technology supports reflecting on my behaviors as an ongoing activity. & 4.06 (0.90) \\
RUM1 & This technology can put me in a negative thought cycle. & 2 (0.94) \\
\rowcolor{tablepurple}
RUM2 & This technology makes me more likely to ruminate about a past situation. & 2.58 (1.41) \\
RUM3 & Using this technology can make me ruminate or dwell over things that happened for a long time afterward. & 2 (1.06) \\
\rowcolor{tablepurple}
THK1 & This technology makes me feel that it is important to me to understand what my feelings mean. & 3.24 (0.83) \\
THK2 & This technology supports me in usually knowing why I feel the way I do. & 3 (1.12) \\
\rowcolor{tablepurple}
THK3 & This technology makes me feel that it is important to me to be able to understand how my thoughts arise. & 3.59 (0.94) \\ \bottomrule
\end{tabular}
\caption{Summary statistics from the R2T2 scale \cite{Loerakker_Niess_Wozniak_2024} administered to participants in Case 2. The Likert scale questions were presented on a scale of 1 = strongly disagree to 5 = strongly agree. REF = reflection subscale, RUM = rumination subscale, and THK self-focused thinking subscale.}
\label{r2t2}
\end{table}

However, forcing critical distance in the reflection-in-action phase was not always positive for participants; as summed up by P6-C2: "I was less immersed and more focused on thinking about the game critically, which for me was both positive and negative." P1-C1 believed that the documenting "takes away from the emotional impact...[they] don't get as invested because [they're] focusing more on the events and the impacts on the characters, but not how [they] feel about it". Correspondingly, participants discussed their perception that certain games were more appropriate to annotate than others. P2-C2 explicitly mentioned "pick[ing] the games [they] played based on what [they] felt would be good for annotating". P4-C1 and P5-C1 argued that documentation was most valuable in story-rich contexts to "retain key knowledge easier" (P4) or to preserve information that "may be needed at a later time" (P5). Ultimately, as P6-C1 noted, the decision to document "depends on the game and the person". While some players might avoid documenting a fast-paced game like \textit{Hollow Knight} to preserve flow, others might find documenting the same game essential for "figuring out the lore or story" (P6-C1).

These statements lead us to believe that the breaks in immersion can be primarily attributed to the act of reflecting itself, rather than using the system; participants from Case 2 gave SavePoint a mean SUS score of 70.6 $\pm$ 15.63, indicating above average usability and considered acceptable. Taken together, these findings appear to validate \citet{Khaled_2018}'s claim that immersion may "work against enabling us to consider our play experiences - in the moment at least - from an analytical perspective and critical distance". While this is a success for the method's stated goals, it comes with tradeoffs for participants.

\section{Discussion}
In our findings, we illustrate the affordances of using SAVEstate, which included (1) allowing us to connect "the immediate phenomenology of gameplay" \cite{antognozziludic} to reflection-on-action and post-game discourse, (2) understanding how multiple participants reflected-in-action on the same parts of the game, (3) enabling the collection of a wide range of annotation types, and (4) observing meaning-making and endo-game reflection. We identified tensions concerning the Heisenberg Effect \cite{czerwinski2004diary} and how documenting can affect the player experience. In the following sections, we discuss the potential benefits and considerations of using SAVEstate to conduct player-computer interaction research and provide use cases for future study. 

\subsection{Towards Formalizing the Study of Replay}
SAVEstate demonstrated initial promise in capturing ``moment-to-moment player experience'', primarily endo-game meaning-making (Cases 1 and 2), replay (Case 2), and scaffolded exo-game reflection (Case 3) \cite{Whitby_Iacovides_Deterding_2023}. We also observed that some participants found value in stopping to reflect while playing, and that SavePoint successfully supported reflection and self-focused thinking. In supporting how ``[reflection-on-action] may feed back into play and affect subsequent reflection-in-action'', we saw that players in Case 1 - a longer term study which required reengagement - appreciated seeing how their thinking evolved. Taken together, we believe that these findings serve as preliminary evidence for SAVEstate supporting the study of replay and its underlying recursive cycles \cite{antognozziludic, newman2026documentation}. Most GUR methods provide snapshots of play but lack mechanisms for revisiting concrete evidence situated within the game context. Using SAVEstate creates \citet{kay1977personal}'s metaphorical file cabinet for players and researchers alike, from which both parties can easily refer back to specific documents containing concrete evidence of reflection. The documents can also be socially interpreted among other players, and the contextual anchors can be used to ``make new connections and create new pathways'' \cite{marshall1998toward} as players edit and reengage with them over time. Using the framing of \citet{newman2026documentation}'s mechanisms, SAVEstate enabled players to \textit{materially inscribe} evidence of their \textit{interpretive attention}. In Case 2, we observed players \textit{recursively produce} new interpretations of their play through replaying games that were familiar to them. In Case 3, we witnessed how players \textit{socially validated} (or invalidated) the evidence generated by their peers. 

However, we note that more investigation into how players and researchers can jointly negotiate the meaning of SAVEstate documents may be necessary. We observed that the documents carried different value to players than they did to us; while we viewed them as stabilized reflection logs, participants desired to use them as references for placemarkings and problem-working (\citet{marshall1997annotation}'s functions 2 and 4). Participants also found the documents useful for tracking playtime, which was an unexpected byproduct of the method. Future SAVEstate tools must strive to better align the co-construction process such that document utility serves participant needs and researcher objectives alike.

\subsection{Scaffolding and the Reflective Experience}
Through adapting \citet{marshall1997annotation}'s functions of annotations to a game context (see Table \ref{table:combined-annotations}), we observed that the scaffolding and context of the cases led to different proportions of each annotation type. For example, as Case 1 was conducted as an educational activity, there were relatively few instances of player affect (function 5) or incidental reflections (function 6); participants did not give many signals as to their affect (further evidenced by the extremely low emote usage; see Table \ref{table:case-studies}), and primarily annotated on happenings in the game rather than their personal circumstances. As Case 2 encompassed a wide variety of games, we saw evidence for all six function types. Case 3, conducted in a more formal educational space, exclusively saw annotations as interpretive activity (function 3) and as representative of player affect (function 5). As they were explicitly instructed to annotate interpretations, instances of function 3 were unsurprising, but participants frequently utilized the emote function to react to moments in the game that they found amusing (Figure \ref{case3_funny}). 

These observations raised a point regarding the co-construction of player-researcher evidentiary value and the amount of provided scaffolding. While a low degree of scaffolding allowed players to annotate more naturally, the resulting annotations were often difficult to parse (see Section \ref{tensions1}), it also increased the incidence of annotations tracing player affect and marking incidental reflections (functions 5 and 6). A researcher interested in meaning-making and interpretive activity (function 3) would likely not find annotations from functions 5 and 6 particularly salient to their research questions. Relatedly, while we observed annotation as interpretive activity (function 3) across all three case studies, it is notable that we did not observe exo-game reflection outside of Case 3 (see Section \ref{meaningmaking}), in which participants were provided explicit scaffolding to connect their gameplay to their own lives and broader sociocultural forces. 

These results align with \citet{slovak2017reflective}'s extension of \citet{schon1987educating}'s reflective practicum, which highlighted "the need to carefully scaffold the process of reflection, rather than simply assume that the capability to reflect is a broadly available trait to be ‘triggered’ through data". In particular, Case 3 highlighted that while R2 reflection may be a natural byproduct of gameplay, it can remain internal or not verbalized unless explicitly prompted. Thus, usage of SAVEstate and tools built to support it "can be interpreted as directly re-structuring/shaping the experiences of the learners through tasks or specific ‘tools’ to scaffold reflection" \cite{slovak2017reflective}. As highlighted by the tension between immersion and reflection \cite{Khaled_2018} and discussed by participants in our case studies, playing while reflecting is different than playing without intervention, which implies that the experience is indeed being "restructured/shaped". As discussed in Section \ref{tension2}, this may detract from the emotional experience of the game itself. This is a vital consideration when selecting the degree of scaffolding and level of articulation for a SAVEstate study, as the emotional impact of gameplay also begets reflection \cite{Bopp_Mekler_Opwis_2016} and is clearly crucial to eudaimonic PX \cite{bopp2018odd, daneels2021eudaimonic}. Researchers must balance between collecting data on moment-to-moment reflection and the potential emotional impact of the game being studied. In educational games or game-based interventions, which are not typically designed to leave an emotional impact, using SAVEstate with system-initiated or researcher-guided scaffolding is unlikely to change the player experience significantly and can provide valuable information on reflection for the researcher (e.g. Section \ref{edgame}). In studying emotionally impactful games, researchers would need to give careful thought as to when exactly players should reflect in-game (see Section \ref{transform}). 

\subsection{Use Cases and Considerations for SAVEstate}
For researchers interested in using SAVEstate in their research, we offer the following examples of how to apply it to four research contexts that we believe the method is particularly well-suited for studying. We include specific considerations for the reflection-in-action and reflection-on-action phases.

% should make this scaffolding, articulation, review 
\subsubsection{Longitudinal Studies of Eudaimonic Player Experience} \label{transform}
\begin{itemize}
\item \textbf{Motivation}: Using SAVEstate in longitudinal formats could be valuable for collecting instances of replay and how meaning-making evolves over time. Researchers could conduct a SAVEstate study with adolescents or young adults with specific games that they enjoy, as prior research has shown that games played in formative periods can spark meaningful effects later in life. Having participants play a game that they have a pre-existing relationship with may be preferable to ensure that the emotional impact of the first playthrough is not affected by the act of documenting. 
\item \textbf{Degree of scaffolding}: Researchers may be the most interested in recording interpretive activity (Function 3), which could be prompted through \textit{medium} scaffolding using language similar to \citet{Whitby_Iacovides_Deterding_2023}: "Observe when you find that the game tries to challenge the way you think or feel about something." Alternatively, researchers could implement \textit{high} scaffolding where the system prompts the player during in-game moments which would be considered one of \citet{Miller_Gandhi_Whitby_Kosa_Cooper_Mekler_Iacovides_2024}'s \textit{Slowdowns}. For example, it may be appropriate to prompt reflection-in-action during a \textit{vista}, \textit{infinite moment}, or \textit{safe place} design pattern.
\item \textbf{Reflection-in-action phase}: For ease of use over extended periods, researchers should consider allowing for \textit{all} articulation levels.
\item \textbf{Level of review}: The level of review should be at least \textit{medium}, as participants should be encouraged to edit documents so that replay and reengagement can be studied. Depending on the duration of the longitudinal study, researchers could also integrate high levels of review weekly or monthly, giving participants structured prompts asking them to review their documents and record how their thoughts have evolved over time. 
\end{itemize}

In addition to annotating while playing, it may be valuable for participants to annotate while engaging in participatory practices, such as watching YouTube videos or using forums. However, it is imperative for researchers to consider participant privacy implications when using SAVEstate; for example, it would be inappropriate to annotate participation in a private Discord server without informing the other members. 

\subsubsection{Scaffolding Reflection in Educational Games} \label{edgame}
\begin{itemize}
\item \textbf{Motivation}: In educational contexts, where the primary goal is typically to convey specific learning outcomes rather than emotional impact, using SAVEstate can help researchers connect reflections to post-game discourse and identify which parts of the game prompt the most reflection.
\item \textbf{Degree of scaffolding}: Researchers should utilize \textit{high} scaffolding to ensure that contextual anchors are created in alignment with learning objectives.  
\item \textbf{Level of articulation}: Researchers should use a \textit{high} level of articulation to prompt instances of interpretive activity (\citet{marshall1997annotation}'s function 3) and problem-working (\citet{marshall1997annotation}'s function 4).
\item \textbf{Level of review}: A \textit{high} level of review would be appropriate for an educational setting. If participants are playing individually, participants can be prompted with structured questions to provide a reflective summary at the end of their sessions. This can be stored as a useful part of the stabilized document itself. In social classroom settings, participants can additionally engage in instructor-moderated debriefs and utilize their annotations in discussions. 
\end{itemize}

\subsubsection{Game Preservation} \label{gamepres}
\begin{itemize}
    \item \textbf{Motivation}: Game scholars emphasize preserving PX alongside code \cite{newman2011not, nylund2015walkthrough, barwick2011playing, swalwell2017fans}. While emulation can capture some aspects of PX, it is less capable of capturing in-situ reflection or evolution of subjective interpretations. SAVEstate offers a supplement to other player generated content for game preservation such as walkthroughs \cite{nylund2015walkthrough, newman2011not, newman2024participatory}. This can capture both personal and community-based interpretations of the game object. In this case, it is important for preservationists to consider what knowledge of the game they would like to preserve, as this will inform the selection of appropriate scaffolding. To demonstrate this, we offer suggestions for capturing experiential knowledge (gained from playing the game) and participatory knowledge (gained from engaging with the game community) \cite{newman2024participatory}. 
    \item \textbf{Degree of scaffolding}: Researchers interested in preserving experiential knowledge should use \textit{medium} scaffolding if there is a specific aspect of play they want to capture (e.g., reasoning of why players make a choice in the game). In contrast, if the goal is to produce examples of community driven usages, \textit{high} scaffolding could prompt system-initiated reflection aligning with well-established community norms (e.g., a common glitch). This could demonstrate how players bring in community knowledge and personal experiences into the game environments. 
    \item \textbf{Level of articulation}: A \textit{high} level of articulation would likely fit this research question, as written text may integrate best into pre-established systems of knowledge organization.
    \item \textbf{Level of review}: Capturing how games and their meanings change over time is central to preservation \cite{barwick2011playing}. A \textit{medium} level of review allows players or preservationists to add, delete, or re-interpret anchors, treating these revisions as longitudinal evidence. Each iteration of the annotated record serves as a distinct document of the game’s evolving state at different points in time.
\end{itemize}
%Michele - game preservation section

% \subsubsection{Playtesting}
% % playtesting
% % for indie devs who maybe don't have that much access to user research infra (check book), SAVEstate could maybe replace think aloud. high scaffolding wtih eye towards usability issues, player thoughts, etc. definitely audio
% \begin{itemize}
%     \item \textbf{Motivation}: For independent developers with limited access to traditional user research infrastructure, using SAVEstate could be a lightweight alternative to think-aloud protocols.
%     \item \textbf{Reflection-in-action phase}: This phase should utilize high scaffolding with a focus on capturing usability issues and systemic friction points. Audio input may be the ideal modality for articulating \textbf{procedural signals} (Function 1) and \textbf{problem-working} (Function 4).
%     \item \textbf{Reflection-on-action phase}: 
% \end{itemize}

\section{Limitations}
\subsection{Limits of the Method}
SAVEstate has several limitations as a method. As previously discussed, its purposeful creation of critical distance may detract from emotional impact, likely making it inappropriate for first playthroughs of games thought to evoke eudaimonic PX. It is also inappropriate for fast-paced multiplayer genres such as first-person shooters or MOBAs, as these games do not generally build in slowdowns \cite{Miller_Gandhi_Whitby_Kosa_Cooper_Mekler_Iacovides_2024}, and provide players with limited opportunities to pause and reflect. Individuals vary in trait self-reflection \cite{silvia2011evaluating}, and the ability to reflect is a skill \cite{fullerton2025well}, meaning that some participants may need more scaffolding than others. Despite designing to keep user interaction natural, some participants still reported feeling observed; future iterations on SavePoint should potentially disable automatic uploads and let players edit sessions before submitting to the research team. We observed limitations when using SAVEstate socially in that group dynamics affected participation in Case 3; shy students not controlling the computer rarely made annotations. 

\subsection{Study Limitations}
In Case 1, technical difficulties with automated uploads initially led to inaccessible post-play reflection screens for three participants. While we patched SavePoint to use local fallbacks so that participants could continue engaging in weekly discussions, we experienced some data loss as we were unable to access SAVEstate documents from participants' local computers. Participants were also limited in geographic and demographic diversity. This was particularly true for Case 2, where many participants were games researchers interested in using SavePoint for their own projects. Our longest case study took place over a month, which is perhaps reflective of a meso-loop of meaning-making but not a macro-loop; longitudinal studies that use SAVEstate over several years are necessary for full maturity of the method. The original definition of SAVEstate did not allow for oral reflections, and therefore, we did not deeply investigate its usage outside of Case 3. We also did not investigate how participants might interface with system-initiated scaffolding. 

\section{Conclusion \& Future Work}
% comparing audio and text
% doing long term study
In this work, we posited that engaging in documentary practices in-situ to gameplay would create the critical distance necessary for reflection-on-action, while additionally providing researchers and players with valuable artifacts from micro-loops of play. To this end, we developed the SAVEstate method and an accompanying data collection tool called SavePoint to probe players' reflection-in-action and how it affected subsequent reflection-on-action. Using SavePoint, we conducted three case studies based on the SAVEstate's primary affordances, namely (1) prolonged engagement with a single game, (2) in-situ reflection in natural environments, and (3) scaffolding dialogical reflection. We were able to observe in-situ meaning-making and connect it to post-game reflection-on-action, as well as synchronous sensemaking across multiple participants. We also identified how annotations functioned within the context of gameplay. We discussed implications using SAVEstate to study eudaimonic PX, educational and serious games, and game preservation. 

In future work, we intend to investigate the effect of scaffolding in SAVEstate using controlled experiments, and how AI might be integrated into SavePoint to support structured reflection-on-action. We plan to use SAVEstate to conduct longer term studies (3-6 months) in which we can watch how player reflection and replay in specific games evolves over time. SavePoint can also be used as a collection tool for supplementing other methods, autoethnography (e.g. \cite{vakeva2024disorientation}),  systematic self-observation (e.g. \cite{Whitby_Iacovides_Deterding_2023}), and diary studies (e.g. \cite{Munteanu_Mo_Potapov_George_Miller_Singh_2026}). Additionally, we wish to explore how SavePoint may benefit players seeking to document their own player experience independent of research, and how indie game developers could utilize SavePoint in playtesting. 

In sum, we make methodological and artifact contributions that can be easily and rapidly adapted for studying eudaimonic player experiences, and hope to advance the field's overarching research agenda on how games can have meaningful impacts on people's lives. 

\begin{acks}
Funding for this work was provided in part by the Jacob's Foundation CERES Network. We would like to thank Travis Windleharth for facilitating access to foundry10's youth recruitment channels and Raeesah Azam for assisting our cohort session. The first author would also like to thank Alicia Guo for many valuable conversations, and Rebecca Turner for making several improvements to SavePoint. We are grateful to our anonymous reviewers for their helpful feedback. Finally, we would like to thank all of our participants for their constructive comments. 
\end{acks}

\bibliographystyle{ACM-Reference-Format}
\bibliography{sample-base}
\bibliographystylegame{ACM-Reference-Format}
\bibliographygame{games}
\end{document}